\documentclass[fleqn,usenatbib]{mnras}

\usepackage{newtxtext,newtxmath}

\usepackage[T1]{fontenc}

\DeclareRobustCommand{\VAN}[3]{#2}
\let\VANthebibliography\thebibliography
\def\thebibliography{\DeclareRobustCommand{\VAN}[3]{##3}\VANthebibliography}

\usepackage{graphicx}	
\usepackage{amsmath}	
\usepackage[table]{xcolor}
\usepackage{caption,subcaption}
\usepackage{ulem}

\newcommand{\tth}{The300 }
\newcommand{\rvir}{$R_{200c}$}
\newcommand{\mvir}{$M_{200c}$}
\newcommand{\ir}[1]{#1}
\newcommand{\rr}[1]{\textcolor{black}{ #1}}

\title[BCGs Trace Mass Bias and Triaxiality]{Brightest Cluster Galaxies Trace Weak Lensing Mass Bias and Halo Triaxiality \ir{in The Three Hundred Project}}

\author[]{
Ricardo~Herbonnet$^{1}$\thanks{E-mail:ricardo.herbonnet@stonybrook.edu}, 
Adrian~Crawford$^{2,3}$\thanks{E-mail:adrian.crawford@virginia.edu},
Camille~Avestruz$^{2,4}$\thanks{E-mail:cavestru@umich.edu},
Elena~Rasia$^{5,6}$, 
Carlo~Giocoli$^{7,8}$,
\newauthor
Massimo~Meneghetti$^{7}$,
Anja von der Linden$^{1}$,
\ir{Weiguang Cui$^{9}$},
\ir{Gustavo Yepes$^{10,11}$}
\\
$^{1}$Department of Physics and Astronomy, Stony Brook University, Stony Brook University, Stony Brook, NY 11794, USA\\
$^{2}$Department of Physics, University of Michigan, 450 Church St, Ann Arbor, MI 48109, USA\\
$^{3}$Department of Astronomy, University of Virginia, 530 McCormick Rd, Charlottesville, VA 22904, USA\\
$^{4}$Leinweber Center for Theoretical Physics, University of Michigan, 450 Church St, Ann Arbor, MI 48109, USA \\
Via Gobetti 93/3, I-40129, Bologna, Italy\\
$^{5}$National Institute for Astrophysics, Astronomical Observatory of Trieste (INAF-OATs), Via Tiepolo 11, 34131 Trieste, Italy \\
$^{6}$Institute for Fundamental Physics of the Universe (IFPU), Via Beirut 2, 34014 Trieste, Italy \\
$^{7}$INAF - Osservatorio di Astrofisica e Scienza dello Spazio di Bologna, via Gobetti 93/3, I-40129 Bologna, Italy\\
$^{8}$INFN - Sezione di Bologna, viale Berti Pichat 6/2, I-40127 Bologna, Italy \\
\ir{$^{9}$Institute for Astronomy, University of Edinburgh,  Blackford Hill, Edinburgh, EH9 3HJ,  United Kingdom}\\
\ir{$^{10}$Departamento de Fısica Teorica, Modulo 8, Facultad de Ciencias, Universidad Autonoma de Madrid, 28049 Madrid, Spain\\}
\ir{$^{11}$CIAFF, Facultad de Ciencias, Universidad Autonoma de Madrid, 28049 Madrid, Spain}
}
\date{Accepted XXX. Received YYY; in original form ZZZ}

\pubyear{2021}

\begin{document}
\label{firstpage}
\pagerange{\pageref{firstpage}--\pageref{lastpage}}
\maketitle

\begin{abstract}
Galaxy clusters have a triaxial matter distribution.  The weak-lensing signal, an important part in cosmological studies, measures the projected mass of all matter along the line-of-sight, and therefore changes with the orientation of the cluster.
Studies suggest that the shape of the brightest cluster galaxy (BCG) in the centre of the cluster traces the underlying halo shape, enabling a method to account for projection effects. 
We use 324 simulated clusters at four redshifts between 0.1 and 0.6 from `The Three Hundred Project' to quantify correlations between the orientation and shape of the BCG and the halo. 
We find that haloes and their embedded BCGs are aligned, with an average $\sim$20 degree angle between their major axes. \ir{The bias in w}eak lensing cluster mass \ir{estimates} correlate\ir{s} with the orientation of both the halo and the BCG. \ir{Mimicking observations, we compute} the projected shape of the BCG, as a measure of the BCG orientation, and find that it is most strongly correlated to the weak-lensing mass for relaxed clusters. We also test a 2-dimensional \ir{cluster} relaxation proxy measured from BCG mass isocontours.
The concentration of stellar mass in the projected BCG core compared to the total stellar mass provides an alternative proxy for the BCG orientation. We find that the concentration does not correlate to the weak-lensing mass bias, but \ir{does correlate with} the true halo mass.
These results indicate that the BCG \ir{shape and orientation} for large samples of relaxed clusters can provide information to improve weak-lensing mass estimates. \end{abstract}




\section{Introduction}\label{sec:intro}
Galaxy clusters are rare objects, known as the largest virialized objects in the Universe, which, according to the current cosmological model, have formed through the hierarchical merging of smaller dark matter haloes. This merger scheme predicts the number of haloes of a given mass (halo mass function) for a given cosmology. An observational census of haloes thus provides a cosmological probe \citep[e.g.][]{sheth99, despali16}. At the high mass end, the halo mass function is steep, meaning that clusters-size haloes have great leverage over the normalisation of the mass function and are therefore powerful cosmological probes \citep{vikhlinin09,  mantz15, dodelson16, bocquet19, to21}. For a review on cluster cosmology see e.g. \citet{allen11}. 

A galaxy cluster census aims to determine both the number of clusters and their total masses to constrain cosmological parameters. 
Clusters can be \ir{detected} in optical \citep{rykoff16, maturi19, aguena21}, millimeter \citep{bleem20,hilton21} and X-ray observations \citep{vikhlinin09a,liu21}.
Halo masses are usually determined from the baryonic observables used to detect the clusters, but these have to be calibrated using unbiased mass estimators. Weak gravitational lensing has become the standard method for this correction \citep[e.g.][]{vdlinden14, mcclintock19, herbonnet20, schrabback20, descl20, lesci20}. The gravitational potential of the cluster introduces a coherent distortion in the observed shapes of galaxies behind the cluster, which is directly related to the mass of the cluster. 
Weak lensing is sensitive to all matter along the line-of-sight and thus measures the total projected mass. To relate this to the spherical overdensity masses of the halo mass function, spherical symmetry of the haloes is incorrectly assumed. Simulations have shown that weak lensing cluster masses are almost unbiased, but the random orientation of the cluster's triaxial mass distribution introduces a $\sim$20 scatter (orientation bias), as well as contributions from large scale structure along the line-of-sight \citep[e.g.][]{meneghetti10,becker11,giocoli14,meneghetti14}. 

The scatter due to projection effects in the weak lensing mass can be mitigated by using large samples, but only when the cluster detection is unaffected by projection. This is not the case for optical cluster finders \citep{dietrich14, sunayama20} and to a lesser extent also for millimeter cluster finders \citep{shirasaki16}. Upcoming millimeter and optical surveys  are projected to find tens of thousands of galaxies over almost a hemisphere in the coming decade. Projection effects will need to be addressed in order to reliably infer cosmology with galaxy clusters. 

One way to deal with projection effects is to model \rr{their} effect on the relation between cluster observable and halo mass \citep[e.g.][]{ costanzi20}. However, a practical estimator of the dark matter halo orientation could provide a way to select cluster samples truly representative of the whole population. 
In simulations of galaxy clusters it has been shown that central galaxies in clusters, also known as the brightest cluster galaxies (BCGs), grow through mergers with satellite galaxies, where the merger timescale scales inversely with satellite mass. Therefore, central galaxies mainly merge with other central galaxies when their parent haloes merge \citep{delucia07}. This manner of growth implies that BCGs accrete matter along the same infall direction as the parent halo, and the mass distributions of BCG and halo should have the same orientation. Indeed, central galaxies have been shown to be aligned with their cluster halo in simulations \citep[e.g.][]{ragone20,depropris21} and observations \citep[e.g][]{donahue16,durret19,wittman19}. The extended envelope of stars around the BCG, called the intracluster light, is also a good tracer of the dark matter distribution \citep{montes19}. \rr{Multi-wavelength observations have also indicated that galaxy clusters exhibit an alignment between the BCG, gas (from X-ray and millimeter), and weak lensing signatures \cite{donahue16}.}
Based on \ir{the expected alignment between the BCG and cluster halo}, several observational studies have shown that the observed shape of the BCG, as a proxy for the orientation with respect to the line-of-sight, correlates to weak lensing mass \citep{marrone12,mahdavi13,gruen14, herbonnet19}.

In this paper we investigate in detail the correlation between the shape \ir{and orientation} of the BCG and \ir{those of} the cluster halo in simulations, where both 3D orientations of mass distribution and the projected 2D shapes can be measured. We use the clusters from \tth project, which has simulated 324 clusters with full hydrodynamical physics \citep{cui18}.
The large sample of clusters available in \tth is particularly important for our study. First, this is required for a precise measurement of the scatter in the weak lensing mass. Second, measuring shapes of objects is non-trivial, due to the proximity of nearby objects and the difficulty in establishing concrete boundaries to the extent of an object (see also Figure~\ref{fig:bcg_shapes}). With a large sample we can remove objects with very uncertain shape measurements without affecting our results too much.

In Section~\ref{sec:methods} we describe our data and methods, in Section~\ref{sec:3dresults} we look at the alignment between the BCG and cluster halo in three dimensions and how the orientation of both relates to weak-lensing mass measurements. We look at projected quantities of clusters in Section~\ref{sec:2dresults}, as this is what can be observed in the real Universe. In Section~\ref{sec:massdistr} we look at an alternative method to estimate the orientation of the BCG and we conclude in Section~\ref{sec:conclu}.


\section{Methodology}\label{sec:methods}

In this section we present the data we use and the analyses we perform. 

\subsection{Data: \tth Project}\label{sec:300}


\subsubsection{General: The Simulated Sample}

We use the most massive galaxy clusters found in the zoom-in simulated regions from \tth\footnote{https://the300-project.org}.
\citet{cui18} fully details \tth Project, but we briefly describe the simulated sample here.  
Our sample is extracted from 324 regions built around the most massive clusters identified at $z$=0 in the dark-matter-only MultiDark simulation \citep{klypin16}, specifically the Planck2 box. The parent simulation consists of a box with sides of co-moving length 1 $h^{-1}$ Gpc, and contains 3840$^3$ particles each of mass $1.5 \times 10^9 M_\odot$. The Planck2 box uses cosmological parameters from \citet{planckcosmo16} ($\Omega_m$ = 0.307, $\Omega_b$ = 0.048, $\Omega_\Lambda$ = 0.693, h = 0.678, $\sigma_8$ = 0.823, $n_s$ = 0.96). 

\tth consists of the zoom-\rr{in} resimulations of these 324 Lagrangian regions including full baryon physics.  The mass range of \tth spans $6.4\cdot10^{14}~M_\odot<M_{200c}<26.5 \cdot 10^{14} M_\odot$ at $z=0$, where $M_{200c}$ is the mass within a cluster-centric sphere of radius $R_{200c}$ enclosing an average density that is 200 times the critical density of the universe. The resimulation of each cluster includes high-resolution particles within a spherical region of radius 15 $h^{-1}$ Mpc at z = 0, centred on the highest density peak \ir{of the main cluster}.  For the resimulation, the respective dark matter and gas particle masses are $m_\mathrm{DM} = 12.7 \times 10^8 h^{-1} M_\odot$ and $m_\mathrm{gas} = 2.36 \times 10^8 h^{-1} M_\odot$. The simulations have dark matter Plummer smoothing length of 6.5~kpc/h. Outside the high-resolution regions, only dark matter particles at lower resolution are kept to properly trace the large-scale gravitational field.

\ir{\tth Project includes resimulations with three different versions of hydrodynamic simulation codes with baryon models: Gadget-MUSIC, Gadget-X, and GIZMO-SIMBA \citep{cui22}. For this analysis, we use the resimulations generated with the smoothed-particle hydrodynamics scheme and baryonic implementations in the full physics Gadget-X code \citep{rasia15,beck16}.} The dataset consists of 128 simulation snapshots saved between $0\leq z\leq 17$, and \ir{halo catalogs} from the Amiga Halofinder \citep{knollmannandknebe09}. 
Even if the Lagrangian regions are large enough to contain other massive clusters, we only consider the most massive object in each region, with the exception of the few clusters that, at the considered redshift, were contaminated by low-resolution particles. In this work, we primarily use data from four snapshots \ir{summarized in Table~\ref{tab:cl_samples}}, $z=0.116$ \ir{(snapshot 123)}, $z=0.220$ \ir{(119)}, $z=0.333$ \ir{(115)}, and $z=0.592$ \ir{(107)}. 
These redshifts are roughly representative of the range found in cluster weak-lensing analyses \citep[e.g.][]{descl20, giocoli21}. Some studies target more distant clusters \citep[e.g.][]{chiu20,schrabback20}, but for our redshifts we can be sure that the shape of the BCG could be reliably measured in observations.

\begin{table}
    \centering
    \begin{tabular}{c|c|c|c}
    Snapshot & Redshift & N$_{\text{3d,clusters}}$ & M$_\mathrm{200,min/max}$\\
    \hline
    123    & 0.116 & 320 (324) & [$23.0\times10^{12}$, $2.2\times10^{14}$]\\
    {\bf 119}    & {\bf 0.220} & {\bf 316 (324)} & [$7.7\times 10^{12}$, $ 1.9\times 10^{14}$]\\
    115    & 0.333 & 315 (324) & [$12.0\times10^{12}$, $2.1\times10^{14}$]\\
    107    & 0.592 & 281 (324) & [$9.3\times10^{12}$,$1.6\times10^{14}$]
    \end{tabular}
    \caption{\ir{Summary of simulated galaxy clusters used in this analysis: snapshot number, corresponding redshift, and number of clusters with available 3-dimensional particle information.  We bold our fiducial snapshot, 119, for which we use all 324 clusters with projected map images.} 
    }
    \label{tab:cl_samples}
\end{table}

Our analysis makes use of projected and three-dimensional distribution of the stellar particles near the central regions of clusters associated with the brightest cluster galaxy (see \ref{sec:bcgmaps}), and the projected spatial distribution of \ir{all the} particles in each galaxy cluster (see \ref{sec:convmaps}). 
Previous works have analyzed and validated various components of the simulations, including galaxy properties \citep{wang18}, gas profiles \citep{mostoghiu19,li20}, and the dynamical states of the galaxy cluster sample \citep{capalbo20,deluca20}.

\subsubsection{Data: Dark matter distribution}\label{sec:convmaps}

For all clusters, we have their true total mass $M_{200c}$ computed by summing over all particle species (dark matter, stars, and gas).
The three-dimensional shapes of the total cluster mass distribution, including gas, stars, and dark matter, were computed by \citet{knebe20} and we use their results. We discuss their shape measurement method in Section~\ref{sec:3dshapes}.

Light rays passing by a galaxy cluster have their trajectories deflected due to the curvature of space-time. In this section, we briefly summarize the procedure adopted to derive the lensing properties of the clusters in our sample \citep{meneghetti10,meneghetti14,meneghetti20}. We will give a more detailed description in a forthcoming paper (Meneghetti et al., in prep.). Given the relatively small size of galaxy clusters compared to the typical distances involved in gravitational lensing phenomena, we can assume that the deflection occurs on a plane, called {lens plane}. We begin by choosing an arbitrary axis passing through us (the observer) and perpendicular to the lens plane, and we compute the positions on the sky relative to this axis.  The lens equation relates the intrinsic and apparent angular positions, $\vec\beta=(\beta_1,\beta_2)$ and $\vec\theta=(\theta_1,\theta_2)$, of a distant source lensed by the cluster:
\begin{equation}
\vec\beta = \vec\theta -\vec\alpha(\vec\theta) \;,
\end{equation}
where $\vec\alpha(\vec\theta)$ is the deflection angle at position $\vec\theta$.

Let $\Sigma(\vec\theta)$ be the cluster surface density at position $\vec\theta$, obtained by projecting all particles on the lens plane. We can define the lens { convergence} as
\begin{equation}
\kappa(\vec\theta)=\frac{\Sigma(\vec\theta)}{\Sigma_{cr}} \;,
\end{equation} 
where 
\begin{equation}
\Sigma_{cr} = \frac{c^2}{4\pi G}\frac{D_{S}}{D_LS D_L}
\end{equation}
is the critical surface density, and $D_L$, $D_S$, and $D_{LS}$ are the angular diameter distances between the observer and the lens, the observer and the source, and the lens and the source, respectively. 
We obtain three convergence maps for each simulated cluster by projecting the masses of all particles along the simulation axes $x$, $y$, and $z$. We select the particles within a volume of depth 10 Mpc centered on the cluster center\footnote{The cluster center coincides with the minimum of the cluster gravitational potential well.}, producing maps of $6\times 6$ Mpc. 
\rr{The line-of-sight depth was chosen because it fits within the spherical high-resolution volume. \citet{becker11} have shown that the weak-lensing mass bias and scatter do not change significantly if the depth was increased to 20 Mpc.}

The deflection angle can be expressed in terms of the convergence via a convolution integral:
\begin{equation}
\vec\alpha(\vec\theta) = \frac{1}{\pi}\int d^2\theta' \kappa(\vec\theta')\frac{\vec\theta-\vec\theta'}{|\vec\theta-\vec\theta'|^2} \;.
\end{equation}
Thus, we can derive the components, $\alpha_1$ and $\alpha_2$ of the deflection angle $\vec\alpha(\vec\theta)$ at each position on the lens plane from the equation above using fast-Fourier-Transform methods \citep[e.g.,][]{PresTeukVettFlan92}.  Since these assume periodic boundary conditions, we remove the outer region of $1$ Mpc surrounding the maps to limit numerical errors. Thus, the resulting deflection angle maps have a size of $5\times 5$ Mpc, spatially resolved with $2048\times 2048$ pixels. To avoid shot noise due to particle discreteness, we apply a Gaussian smoothing with FWHM of $\sim 7$kpc to the convergence maps before computing the deflection angles.  

From the maps of the deflection angles, we derive the shear components $\gamma_1$ and $\gamma_2$, defined as
\begin{eqnarray}
\gamma_1 & = &  \frac{1}{2}\left( \frac{\partial \alpha_1}{\partial \theta_1} - \frac{\partial \alpha_2}{\partial \theta_2}\right) \;, \\
\gamma_2 & =& \frac{\partial \alpha_1}{\partial \theta_2} =  \frac{\partial \alpha_2}{\partial \theta_1} \;.
\end{eqnarray}

In the weak lensing regime, convergence and shear at the image position fully describe how the source shape changes because of lensing. For example, circular sources are mapped onto elliptical images, whose major and minor axes have lengths
\begin{eqnarray}
a = \frac{1}{1-\kappa-\gamma} \;, \\
b = \frac{1}{1-\kappa+\gamma} \;.
\end{eqnarray} 
In the formulas above,  $\gamma = \sqrt{\gamma_1^2+\gamma_2^2}$ is the shear modulus. The source magnification is given by $\mu=[(1-\kappa)^2-\gamma^2]^{-1}$. More generally, the measured ellipticity of a lensed source, given by $e=(a-b)/(a+b)$, provides an unbiased estimate of the so-called {reduced shear}, $g=\gamma/(1-\kappa)$.

\rr{We calculate the critical surface density assuming a fiducial source redshift $z_s=3$, noting that weak lensing measurements that incorporate a true redshift distribution would require rescaling of these convergence maps.  These maps provide projected surface density maps from which we derive projected halo shape and mass measurements  (see Section 2.2.1 for more information).}

\subsubsection{Data: Regions Containing the Brightest Cluster Galaxy}\label{sec:bcgmaps}

We analyse both the 3-dimensional stellar particle distribution and the projected stellar density maps from \tth centered around the cluster density peak. This peak is assumed as both the centre of the galaxy cluster and the BCG.

To measure 3-dimensional shapes \rr{of the BCGs }we use all the stellar particles within a sphere of radius 100 $h^{-1}$ kpc around the cluster centre. Each particle has $x,y,z$ coordinates and mass $m$. \ir{The three-dimensional stellar particle} data is only available for \ir{subsets of the 324 resimulated clusters at each redshift:} 316 at our fiducial snapshot 119\ir{ at $z=0.220$}, 320 at $z=0.116$, 315 at $z=0.333$, and 281 at $z=0.592$\ir{, specified in Table~\ref{tab:cl_samples}.}  

The projected stellar density maps were constructed from a cube of 0.4 Mpc on a side, centred on the density peak, projected along the three main axes of the cube. Similar to the weak-lensing maps, these maps were also smoothed, using a Gaussian with FWHM of $\sim$2 kpc. The projected stellar mass maps are 0.4$\times$0.4 Mpc$^2$, larger than the 3D data we used, allowing us to study the outer envelope of the BCG. There are \ir{three projections for each of the 324 resimulated clusters totaling to 972 projected stellar density maps at each redshift.}

There are two caveats to our analysis using the simulated data for direct comparisons with observations.  The first is in the properties of simulated brightest cluster galaxies, and the second is in the difficulty of distentangling stars associated with the brightest cluster galaxy and the intracluster light.  

First, we note that simulating realistic galaxies is difficult and properties of the simulated BCGs we study do not fully match properties of observed BCGs \citep{cui18}.   Clusters from \tth have central galaxies that are relatively more massive and bluer that those in observations.  The differences come from difficulties in capturing microphysical processes with subgrid models.  For example, the quenching of star formation \ir{at redshift $z=0$} is not accurately reproduced, leading to bluer galaxies in the simulations \citep{cui18}.  Additionally, the projected images contain information of the projected stellar mass, integrated along the line-of-sight.  The images do not correspond to flux or luminosity.  Using the projected stellar density maps as a proxy for observations therefore implicitly assumes a constant mass-to-light ratio.  Since our analysis is mostly concerned with the shape of the stellar light, the shape measurements are likely not heavily impacted with this assumption. 

Second, there is generally not a clear-cut distinction between stellar particles that comprise the BCG and the stellar particles that make up the intra-cluster light (ICL) \ir{in neither simulations nor observations \citep[e.g.][]{cui14}}. \ir{Additionally,} the stellar components associated with the ICL \ir{in observations} is usually difficult to see above the noise because of its low surface brightness and observations more easily pick up the brightest stellar components that comprise the BCG \citep{zhang19}.  While there are some methods to try to disentangle the BCG and ICL with dynamics (with phase space information in simulations, e.g. \citet{canas20}), observations find that the surface brightness indistinguishably embeds the BCG and ICL components \citep{kluge20}.  The most straightforward way to approximate the BCG-ICL separation in simulations that is consistent with what observers might do is to use a radial cut.  We therefore measure the shape of the stellar particle distribution at various radii to quantify the differences in stellar density shapes when the BCG likely encloses more or fewer stars that may be associated with the ICL.  

We also note that our construction of centering the stellar density maps on the cluster density peak assumes that the location of the BCG coincides with this definition of the cluster center.  While this typically holds for galaxy clusters in simulation, this is not always the case in observed galaxy clusters since recent major mergers may displace the BCG, leading to oscillations about the peak of the potential 
\citep{depropris21}.

\subsubsection{Data: Relaxation Criteria for Subselection}\label{sec:relax}

The dynamical state of the clusters of \tth have been studied in a few works \citep{cui18, capalbo20, deluca20,haggar20} and we use here a sub-sample of relaxed objects. Relaxed clusters refer to systems that have not undergone recent major mergers or periods of high accretion that drive components of the galaxy cluster further from dynamical or hydrostatic equilibrium.  These clusters exhibit signatures that tend to correlate with equilibrium, such as in the shape of the overall halo \citep{kasunandevrard05,faltenbacher05} or gas shape \citep{chen19,machado20}, or the offset between X-ray gas centers and the centers of collisionless components such as the peak of the dark matter potential or the BCG \citep{depropris21}.  

However, the state of dynamical relaxedness is \ir{not a binary state, rather on a continuum;} galaxy clusters continually accrete matter through filamentary structures. \ir{Typically, some thresholds of dynamical state indicators are chosen as the criteria for binary classification of ``dynamically relaxed''.}  We use two sets of \ir{relaxation} criteria, based on properties computed within $R_{200c}$, to examine how the alignment between the BCG and halo depends on the \ir{subselected samples}.

The relaxed halos are defined following \citet{deluca20}: (1) the halo's centre of mass is less than 0.1\rvir from the true centre ($\delta x_{\rm halo,CoM} < 0.1$\rvir), and (2) the mass in substructures is less than 10\% of the total halo mass within \rvir ($f_{\rm sub}<0.1$\mvir). 

For comparison we also defined a sample of relaxed clusters following \citet{cui18}, with more stringent criteria: (1) the halo's centre of mass is less than 0.04\rvir from the true centre \rr{($\delta x_{\rm halo,CoM} < 0.04$\rvir)}, (2) the mass in substructures is less than 10\% of the total halo mass within \rvir ($f_{\rm sub}<0.1$\mvir), and (3) the virial ratio $\eta = (2T - E_s)/ |W|$ is $0.85<\eta<1.15$.  Here, T is the total kinetic energy, $E_s$ is the surface pressure energy from both collisionless and gas particles and W is the total potential energy \citep{cui17}. \rr{We use the stricter definition of relaxation from \citet{cui18} as the default definition for relaxed clusters and denote it as relaxed C18. }

\subsection{Measurements}
\subsubsection{Measurements: Weak Lensing Mass Estimates}
We use weak lensing masses computed using the method described in \citet{giocoli21}, and briefly outlined here.
The convergence and shear maps described in Section~\ref{sec:convmaps} were used to construct weak-lensing observables for mock galaxies. The weak-lensing maps show a field of view of 5$\times$5 Mpc$^2$ at the cluster redshift. We populate this field with mock galaxies and fill the field of view with circa 30 source galaxies \rr{per} square arcmin, following a redshift distribution that peaks at $z_S\approx1$. This roughly corresponds to the number of expected galaxies for cluster weak lensing with \textit{Euclid} \rr{and the approximate redshift distribution of}\footnote{http://sci.esa.int/euclid/} observations
\citep{laureijs11}. \rr{Specifically, these simulations have been built using a Euclid-like source redshift distribution constructed using Euclid-like images of clusters from SkyLens \citep{plazas19}, rescaled to 30 galaxies per square arcmin. Note, a similar parametrisation has been adopted in \citet{boldrin12,boldrin16}.  For this analysis, we do not assume any redshift uncertainty for the background galaxy population and randomly assign a position to them in the considered field of view. The convergence and shear maps described in Section~\ref{sec:convmaps} are then rescaled from redshift $z_S=3$ to the considered redshift of the source galaxy.}


The source galaxies are then binned in 24 radial annuli around the
cluster centre from a radius of 0.01 $h^{-1}$ Mpc outwards. Shear
uncertainties include shape noise contribution ($\sigma_\epsilon =
0.3$) and the r.m.s. of the shear profile in the annulus.  The binned tangential shear profiles were fit using the \citet{baltz09} density profile, assuming a truncation radius $r_t$ set to 3 times $R_{200c}$. The mean shear was fit only using bins with more than 10 galaxies.

Mass ($M^\mathrm{WL}_{200c}$) and concentration ($C^\mathrm{WL}$) were free parameters of the model with flat priors from log$_{10}M^\mathrm{WL}_{200c}=$ 13 to 16, and from $C^\mathrm{WL}=1$ to 10, respectively. The model was centred on the known cluster centre, and no miscentring terms were added to the model.  Considering the limited field of view in which the clusters are located, we also neglect the 2-halo term in the modelling function \citep{giocoli21}.  We tested this assumption and found negligible differences in the recovered weak-lensing quantities.

\subsubsection{3D shape measurements}
\label{sec:3dshapes}

Shape measurements of mass distributions in simulations are generally done using the moments of the particle distribution.
The first 3 moments are, 
\begin{align}
    I_0 &= \sum_n m_n w_n  1 , \nonumber \\
    \mathbf{I_1} &= \sum_n m_n w_n  \mathbf{x_n} , \nonumber \\
    \mathbf{I_2} &= \sum_n m_n w_n (\mathbf{x_n} \otimes  \mathbf{x_n}) .
 \label{eq:moments}
\end{align}
Here $\mathbf{x}_n$ is the three dimensional coordinate column vector of the $n$-th particle, and $\mathbf{x}_n \otimes \mathbf{x}_n$ is the outer product of the two coordinate vectors. The origin of the coordinates is the true BCG centre. The mass of the $n$-th particle is $m_n$ and it is assigned a weight $w_n$.
The zeroth order $I_0$ moment gives the total mass; the first order moment vector $\mathbf{I_1}$ can be used to determine the centre of the particle distribution. The eigenvectors $\mathbf{e}$ of the second moment $\mathbf{I_2}$ give the primary axes of the distribution and the square roots of the eigenvalues correspond to $a>b>c$, the axis lengths.

Cluster cores are dense regions; a large number of galaxies are close to each other. \rr{Proximity of another galaxy strongly affects $\mathbf{I_2}$, as large values of $|\mathbf{x}_{n}|$ dominate the contributions to the sum.} 
To mitigate the effect of neighbours, we employ a radial \ir{top-hat} weight function, so that $w_n=1$ if the radial distance to the centre is smaller or equal to $r_\mathrm{lim}$ and $w_n$ is set to 0 beyond the limiting radius. We investigated several limits: $r_\mathrm{lim}$= 25, 50, 75, and 100 $h^{-1}$ kpc. \rr{We find that the choice of radial limit has no significant effect on our results (see Section ~\ref{sec:quant_bcg_halo}).} 

\rr{\citet{knebe20} uses another weight function for their measurement of the shape of the halo, which decreases with radius squared: $w_n = |\mathbf{x}_{n}|^{-2}$. However, this weighting highlights the core of the distribution, which is not necessarily the area of interest for galaxies since astrophysics complicates the galaxy cores often rounding the inner shape. We expect the outskirts of BCGs and the ICL distribution to be more aligned with the halo shape and accretion history of the cluster.  Our selected weighting provides a more flexible approach by looking at different radii of interest.}

In addition, we attempt to flag such instances where a neighbour significantly affects the measured BCG shape. For this, we compute the first moment $\mathbf{I}_1$. All three components of $\mathbf{I}_1$ will be zero if the distribution is completely symmetric around the centre, as we roughly expect for galaxy mass distributions. If the distribution is skewed along axis $i$, for instance due to a neighbouring galaxy, then $I_i$ will be large. We flag BCG shape measurements as contaminated if the norm of $\mathbf{I}_1$ is larger than 0.1 $R_\mathrm{lim}$.

The second moments matrix $\mathbf{I_2}$ from the three dimensional data can also be used to find the 2D shapes for the mass distribution if \rr{its} projected along any of the three axes of the simulation.
By omitting those elements of the 3x3 $\mathbf{I_2}$ matrix corresponding to the axis along which we project, we can construct the 2x2 moments matrix for the projected image.
Note that our star particles are within a sphere and therefore the projection depth is not uniform. More particles will lay on lines-of-sight going through the BCG centre. This may artificially bias these projected 2D shapes to be more circular. The projected images described in Section~\ref{sec:bcgmaps} are a better imitation of observations and we use these to quantify the relation of BCG shape and weak-lensing masses.

The shapes of the total mass distribution of the cluster haloes were determined by \citet{knebe20}. All particles (dark matter, gas and star particles) were used to compute the shapes and they used $w_n=1/r^2_n$, where $r_n$ is the radial distance to the centre of the cluster. This emphasizes the core of the cluster over its outskirts, and we thus expect a stronger correlation of the halo shape to the shape of the BCG than for a shallower weight function.

\begin{figure*}
    \centering
    \includegraphics[scale=0.85]{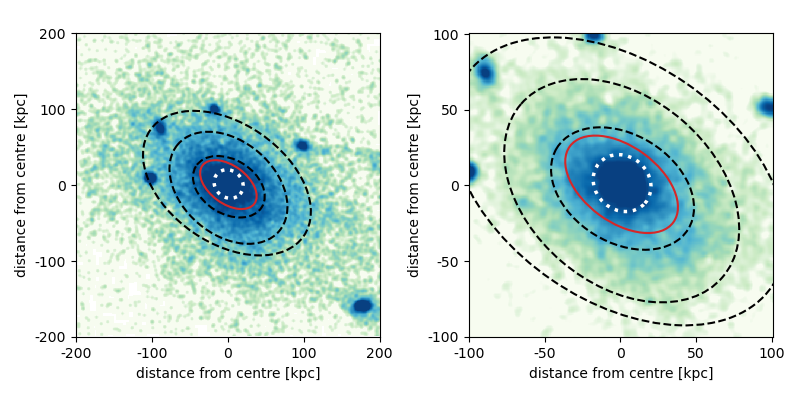}
    \includegraphics[scale=0.85]{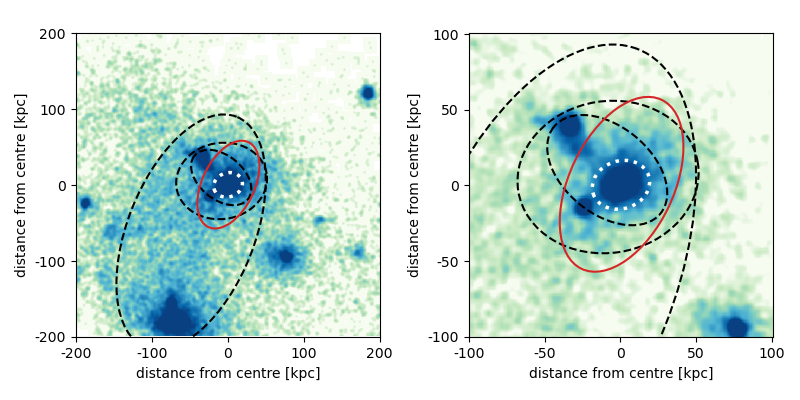}
    \caption{\textit{Top row:} projected stellar density map from which we measured 2D BCG shapes. Full image on the left and zoomed in on the right. The black dashed lines show ellipses drawn using the properties of the contours at 96th, 88th, 80th percentile of the peak brightness in the image, from the image centre outward, respectively. The red solid lines shows the ellipse drawn using the moment measurements. The white dotted line shows the ellipse corresponding to the Sers\'ic model fit at the best-fit half light radius. The Sers\'ic model is a good description of the shape of the fairly round BCG core, but fail to capture the more elliptical envelope of the BCG. The moments prefer a more elliptical shape because they are more sensitive to mass further from the centre. The contours provide a good estimate at the various radii.\\
    \textit{Bottom row:} Same as top panel but showing an example where {\it shape measurements are difficult due to nearby massive objects}. Again the Sers\'ic model describes the core well and are not affected too much by the neighbouring objects. The moments are very affected by the object in the lower left of the image, as is the largest contour. The smaller contours closer to the BCG centre are affected by the nearby objects. \ir{Our flagging routine identifies t}his cluster to have unreliable shape measurements in our analysis.}
    \label{fig:bcg_shapes}
\end{figure*}

\subsubsection{2D shape measurements}\label{sec:2dshapes}

We determine the shape from the projected images with three different methods as a cross-check. 

\noindent \textbf{Isophote contours}: We draw a contour at a fixed isophote using the {\tt python scikit-image} package\footnote{https://scikit-image.org} \citep{walt14}, similar to some observational work \citep{huang18, montes19}. For the stellar density maps, we determine contour shapes at isophotes that correspond to a given percentile brightness with respect to the entire image of the stellar density map, ranging across the percentiles $[80,98]$. Here the 100th percentile corresponds to the densest peak in the map, and for an ideal galaxy density distribution that decreases monotonically with distance from the centre, lower percentile values would trace the shape of the galaxy further out. For a contour, {\tt scikit-image} has properties like its centroid and the lengths of its major axis $a$ and minor axis $b$, from which we compute the axis-ratio $q_{2D}=b/a$ and the radius $r_{2D}=\sqrt{a b}/2$. We enforce that the area enclosed by the contour must include the true BCG centre.
Isophotes at fixed brightness percentile will not be at the \ir{exact} same physical radial distance for each projected cluster. We therefore define the conversion from a brightness percentile to a physical scale as the median of the distribution of radii of all clusters. \ir{The medians range from $55-200$~kpc for the brightest contour to the dimmest contour we measure.  The spread in physical values corresponding to each contour is than 5\% for the innermost contour and 10\% for the outermost contour.}

\noindent \textbf{Image moments:} We compute the image moments for the projected image using Equation~\ref{eq:moments}, where instead of particles of a mass $m_n$ we now have pixels of the convergence map. \rr{The value of the pixel at coordinate $\mathbf{x_n}$ is used instead of $m_n$ in the equation for the second moment.} The eigenvalues of the second moments allow us to determine the 2D axis-ratio $q_{2D}$, similar to the measurements we perform in 3D. We employ a uniform weight function $w_n=1$, emphasizing the contribution of the outskirts of the BCG. 
We compute image moments for both the stellar density and $\kappa$ maps.

\noindent \textbf{Sers\'{i}c profile fits:} 
We fit an elliptical Sers\'{i}c profile to the projected stellar density maps using \texttt{galfit} \citep{peng11}, mimicking some observational work \citep{wittman19,durret19,herbonnet19,zhang19}. The free parameters of the Sers\'{i}c profile are the amplitude, the half-light radius $R_e$ (the radius containing half of the total flux), and the Sers\'{i}c index $n$, as well as the position angle and axis-ratio. We do not mask any parts of the image and use uniform weighting. We found that implementing a mask did not change the resulting shape estimates much on average.\\

In Figure~\ref{fig:bcg_shapes} we show the resulting shape measurements for two different projected stellar mass maps.  \ir{The right panels correspond to the central patch of the left panels, magnified by a factor of 2.} The white ellipse shows a contour of the Sers\'{i}c model, the red ellipse is based on the moments measurements and the black ellipses are based on the properties of the contours at the 80th, 88th, and 96th percentile.
The higher percentile values trace the matter closer to the BCG core. The Sers\'{i}c model traces the core of the BCG, showing that the \texttt{galfit} chi-square minimization is dominated by the dense centre of the BCG.  Moments follow the mass distribution at larger radii, as expected because the weight function is uniform. 

In practice, substructure within the BCG mass distribution or nearby other galaxies will affect the measured shape, as can be seen in the bottom row of Figure~\ref{fig:bcg_shapes}. The \ir{bottom left panel illustrates a case where} shapes of both the largest contour and the moments are dominated by the object in the lower part of the image \ir{$\sim200$~kpc from the center}, which is not part of the BCG.  \ir{The bottom right panel illustrates how} the smaller contours \ir{closer to the centre can be} affected by neighbours. The centroid of the largest black dashed ellipse is notably far from the image centre, where we assume the true BCG is located. In contrast, in the top row of Figure~\ref{fig:bcg_shapes} the contour centroids are very near to the centre. 

\ir{We use centroid offsets to flag instances where the shape measurement is unreliable due to neighbouring objects, such as those illustrated in the bottom panel of ~\ref{fig:bcg_shapes}}.
For a more reproducible flag for observers, who do not know the true BCG centres, we \ir{set} the centroid of a small contour as \ir{a proxy for} the "observed" BCG centroid\ir{, selecting the centroid of the 96th percentile isophote as a default. N}ote that we found little difference when instead using the 98th percentile isophote contour. 
We then compute the offset of the centroid of a larger contour $x_i$ to our "observed" BCG centroid $x_{96}$, where the subscript is labeled by the percentile value of the isophote. 

\ir{We then use the value of the offset normalized by the radius of the larger contour, }$R_i$, which we denote as $\Delta x_{i,96}$.
We found that the distributions of \ir{normalized} centroid offsets is extremely similar regardless of the choice of contours. These distributions all had a long tail, where all shape measurements were heavily affected by neighbouring objects. We only keep shape measurements where $\Delta x_{i,96} < 0.2$, or the centroid \ir{of the contour} is no more than 20\% of its radius away from the "observed" BCG centre. A value of 0.2 cuts off most of the long tail and visual inspection revealed that below 0.2 most shape measurements were fine.
We found that moments and Sers\'ic shape measurements perform similarly to contour shape measurements (see the discussion in Section~\ref{sec:2dshape_comparison}). We therefore \ir{use the same flagging routine} based on contour measurements for the other two methods.

Finally, we also use the contour method \ir{on the convergence maps} to determine the projected shape and orientation of the \ir{dark matter} halo at $R_{200c}$. For this, we compute contours at several isophotes in the convergence maps and use the contour whose radius was closest to $R_{200c}$.

\section{Shape correlations in 3D}\label{sec:3dresults}

In this section, we quantify the alignment between the three dimensional mass distribution of the BCG and the underlying dark matter halo in 324 clusters from 'The Three Hundred Project'.  Then we assess how the alignment relates to the weak lensing mass bias and varies as a function of mass and redshift in our sample. To quantify the correlation between variables, we compute the Spearman correlation with \texttt{pymccorrelation}\footnote{https://github.com/privong/pyMCspearman} and its uncertainty with 1000 bootstraps.  This package implements the Monte Carlo error analysis procedure described in \citet{curran14}.  \citet{privon20} provides analysis with the first use of the package described.

\subsection{Quantifying the BCG-Halo Alignment}\label{sec:quant_bcg_halo}

To determine the BCG-halo alignment from 3-dimensional data, we first compute the second moments for the stellar particles from the region described in Section~\ref{sec:bcgmaps}.  We determine the vector of the major axis $a$ of the stellar particle distribution, $\bf{e}_a^\mathrm{BCG}$, and the corresponding vector for the major axis of the halo, $\bf{e}_a^\mathrm{halo}$.   

The alignment between the halo and the BCG is quantified by the angle between the major axes of both distributions, which we call the (mis)alignment angle, $\alpha$.  We define $\alpha$ as,
\begin{equation}
    \alpha = \mathrm{arccos}( |\bf{e}_a^\mathrm{halo} \cdot \bf{e}_a^\mathrm{BCG} |).
\end{equation}

To quantify effects on projected measurements, we also compute the inclination angle for both the BCG and the dark matter halo distribution.  We define the inclination angle of a mass distribution as the angle between the line-of-sight and the major axis of that mass distribution.  The inclination angle of the BCG is then 
\begin{equation}
    \theta^\mathrm{BCG} = \mathrm{arccos}( |\bf{e}_\mathrm{LOS} \cdot \bf{e}_a^\mathrm{BCG} |),
\end{equation}
where $\mathbf{e}_\mathrm{LOS}$ is the normalised vector along the line-of-sight.  We similarly calculate the inclination angle of the halo.

\begin{figure}
    \centering
    \includegraphics{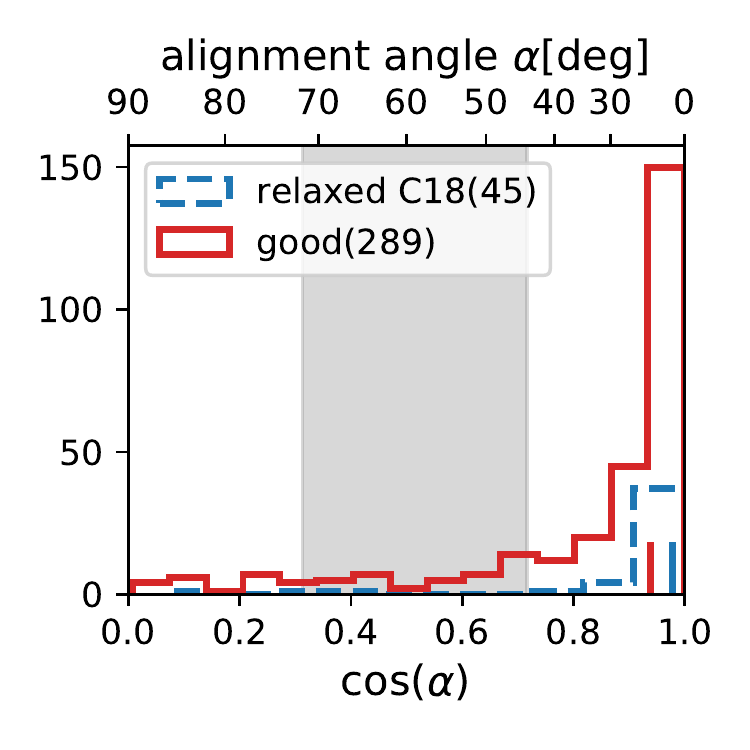}
    \caption{Distribution of the angles between the major axis halo and the major axis of the BGC in degrees and the cosine of this alignment angle for clusters in The300 simulations. \rr{This is shown for a sample of relaxed clusters according to the criteria of \citet{cui18} in blue and for a sample of clusters with BCG shape measurements deemed unbiased by neighbours (good). The number of clusters in each sample is shown in parentheses in the legend.}
    The clusters are taken from a snapshot at $z=0.22$ and a limiting radius of $r_\mathrm{lim}=50$ kpc was used to compute the BCG shape. Short vertical lines (around cos$(\alpha)\sim0.95$) show the median for each sample. The grey shaded area show the 25th to 75th percentile of the distribution of alignment angles for randomly oriented vectors to guide the eye. Both samples have BCGs preferentially aligned with their host haloes.}
    \label{fig:misalignment_hist}
\end{figure}

In Figure \ref{fig:misalignment_hist}, we show the alignment angle $\alpha$ between the major axis of the BCG and the major axis of the halo for the $z\approx0.22$ snapshot (119) for the limiting radius $r_\mathrm{lim}$=50kpc. We show the cosine of the alignment angle because for random orientations of BCG and halo cos($\alpha$) is an approximately flat distribution, whereas $\alpha$ would be highly skewed.
Of the 316 clusters for which we have stellar particle data, we show in red 289 clusters, as 27 clusters were flagged according to the routine described in Section~\ref{sec:3dshapes}. A subselected sample of 45 relaxed clusters according to the criteria from \citet{cui18} is shown in blue. We indicate the median angle of orientation of each population with same-color ticks along the x-axis.

For the entire distribution of unflagged clusters, we find that the orientations of the BCG and the halo are preferentially aligned with a median angle of $\sim$20 degrees ($\cos\alpha=0.94$ denoted with the \rr{red solid line}).  This value is in agreement with other studies \citep{okabe20, ragone20}. For the 45 relaxed, unflagged clusters (\rr{cyan} line) the median alignment angles is lower: 12 degrees. For the \citet{deluca20} relaxation criteria, \ir{which do not have the additional constraint of the virial ratio parameter,} the median angle is $\sim$14 degrees for 85 clusters. 
The decrease in alignment illustrates that BCGs in relaxed clusters tend to be relatively more aligned with their host halo when compared to the entire population. 

Since the orientation of a very spherical or oblate ($a \sim b \geq c$) object is ill-defined, we also looked at a subset of clusters where both the halo and the BCG have $b/a>0.9$. For these 238 clusters we find no significant change compared to the entire unflagged sample.  

We looked at the alignment angle for different choices of the limiting radius $r_\mathrm{lim}$ for the BCG shape measurement and found no significant difference between them. On average, the halo is aligned with the BCG at radii from 25 kpc to 100 kpc. 

\subsection{Weak lensing mass bias}\label{sec:wlbias}

\begin{figure*}
    \centering
    \includegraphics[scale=0.7]{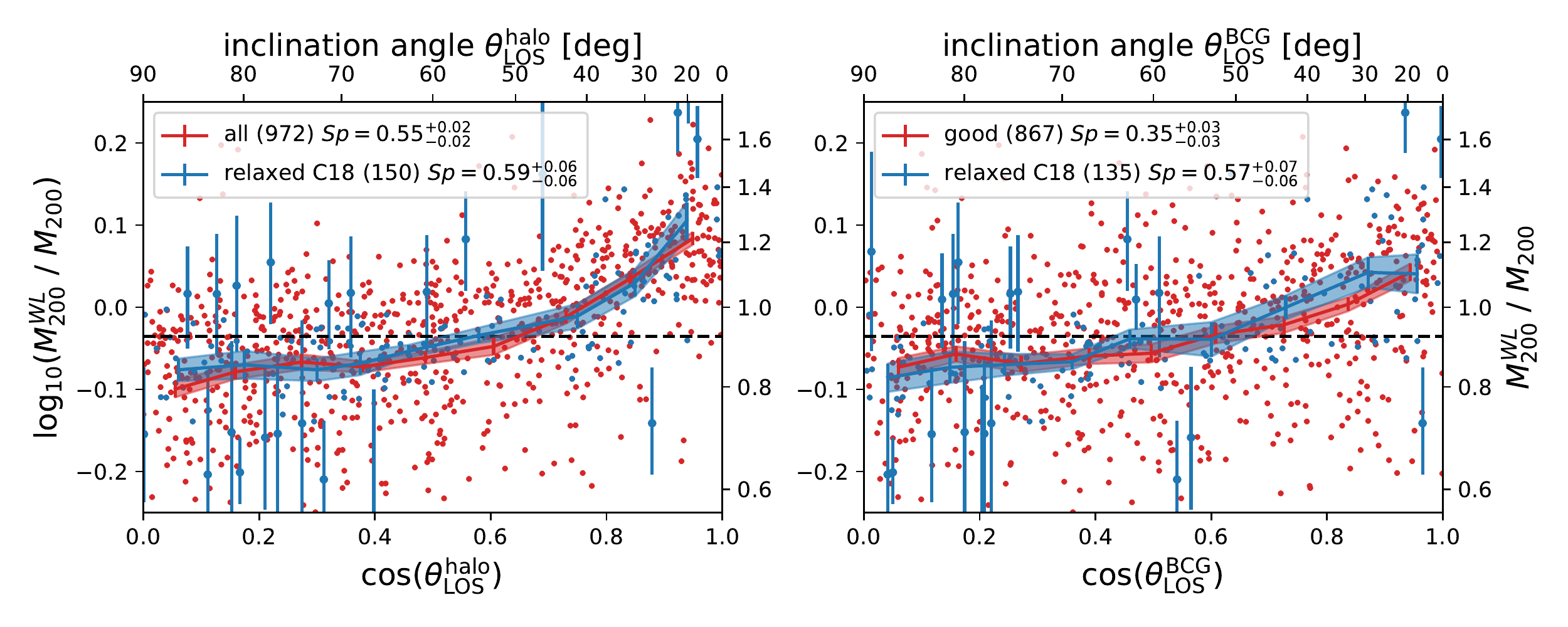}
    \caption{Bias in the weak lensing mass as a function of the inclination angle of the halo particles (\textit{left}) and the BCG stellar particles (\textit{right}) for three projections of clusters at $z\approx0.22$. The mass bias $M^\mathrm{WL}_{200}/M_{200}$ is shown on the righthand y-axis and the log$_{10}$ of the mass bias is shown on the lefthand y-axis for both plots. Clusters oriented along the line of sight are at $\mathrm{cos}(\theta_\mathrm{LOS})\approx1$ and clusters oriented along the plane of the sky at $\mathrm{cos}(\theta_\mathrm{LOS})\approx0$.
    The black dotted line indicates the mean mass bias for all clusters. Coloured lines with errorbars show the mean and uncertainty on the mean in bins of inclination angle for the same selections of clusters as in Figure~\ref{fig:misalignment_hist}, with the number of clusters in each selection noted in the legend in parentheses. There is a clear relation between the inclination angle and mass bias in both panels. As a quantification the legend notes the Spearman correlation $Sp$ for each selection. For clarity, only some \rr{data points} are shown with errorbars coming from the weak-lensing mass estimate. \ir{Note, we limit the range of the y-axis to zoom in on the mean behavior of the trends shown with the bands; }this omits 37 \ir{data points} from the plot. 
    }
    \label{fig:MWLorientation}
\end{figure*}

The assumption of a spherical halo mass profile for the triaxial halo leads to a scatter in the weak lensing mass around the true mass \citep{meneghetti10,meneghetti14}.  We explore this orientation bias by examining the relation of the weak lensing mass bias with inclination angle.
In the left panel of Figure~\ref{fig:MWLorientation} we find for our cluster sample that the direction of the halo major axis is a strong indicator of the bias, in line with results from other studies \citep[e.g.][]{henson17}.
On average clusters masses are over or underestimated by $\sim20$\%, depending on the inclination angle. The mean relation does not change significantly if only relaxed clusters are selected. 
A more detailed analysis of the weak-lensing masses will be described in Giocoli et al. (in prep.). \rr{We find that the scatter in the weak lensing masses about the true cluster mass is 23\% in our simulated sample (20\% for relaxed clusters). On the other hand, the scatter about the mean relation between halo mass and halo inclination (shown as the coloured lines in Figure~\ref{fig:MWLorientation}) is 15\% (11\% for relaxed clusters), a relatively tighter scatter. This shows that halo orientation information can lead to more precise weak-lensing mass estimates.}

\rr{From Figure~\ref{fig:misalignment_hist}, we see that} the BCG and halo major are preferentially aligned. We now investigate the correlation between the BCG and weak lensing mass \ir{bias}. In the right panel of Figure~\ref{fig:MWLorientation}, we again see a clear trend between the inclination angle and bias.  However, the relation is shallower than for the halo shape, due to the imperfect alignment between halo and BCG. The two cluster samples have similar mean relations, but the relaxed sample shows a stronger correlation with a Spearman correlation coefficient $Sp=0.57$, a similar value to the correlation between halo orientation and mass bias. This is likely due to the stronger alignment between halo and BCG for relaxed clusters. \rr{The scatter about the mean relation between halo mass and BCG inclination is 17\% (11\% for relaxed clusters).  This now highlights the potential how BCG information might mitigate the statistical uncertainties in cluster masses, which again is 23\% (20\% for relaxed clusters) in our sample.}

\subsection{Variation with cluster mass and redshift}

There is evidence that the mass and the elongation of a cluster halo are correlated, with lower mass clusters being on average more spherical \citep{despali14,henson17,okabe20}.
We looked at the axis-ratio $b/a$ of the halo and the BCG as a function of redshift and mass. In agreement with these works, we find that haloes become more elongated for higher masses, although the effect is very small, consistent with no trend within the uncertainties. \rr{Over the full mass range the mean axis-ratio $\langle q_\mathrm{halo} \rangle =0.80 \pm 0.08$ decreases by only $\sim$0.04. BCGs are more elongated on average than their haloes with an axis-ratio $\langle q_\mathrm{BCG} \rangle =0.74 \pm 0.10$, but get slightly rounder with increasing cluster mass: $\langle q_\mathrm{BCG} \rangle$ increases $\sim$0.04.}
Similar trends are seen for both haloes and BCGs for clusters at redshifts $z=0.116, 0.333, 0.592$.

To check if our results depend on cluster mass we divided the clusters in to four bins with approximately equal number of clusters. We find that the weak lensing mass bias is lower for the least massive clusters, and the other mass bins are consistent with each other and the full sample. However, the relations between weak-lensing mass and inclination angle for both the halo and BCG are qualitatively the same for all mass bins: the weak lensing mass bias increases when the inclination angle decreases from 90 degrees to 0 degrees. We do not see a significantly shallower slope or different Spearman correlation strength for the lowest mass bin.  \ir{Note, massive clusters still comprise the lowest mass bin, as there are very few groups in the simulated sample.}
 
We repeated our analysis for other snapshots of the simulated cluster regions, at redshifts $z=0.116, 0.333, 0.592$, in addition to our analysis at the fiducial $z=0.221$. We find no significant change with redshift in the relations shown in Figure~\ref{fig:MWLorientation}.

\section{Projected shape correlations}\label{sec:2dresults}

\subsection{BCG inclination and projected shape relationship}
\label{sec:inclination_q}

\begin{figure*}
    \centering
    \includegraphics[scale=0.85]{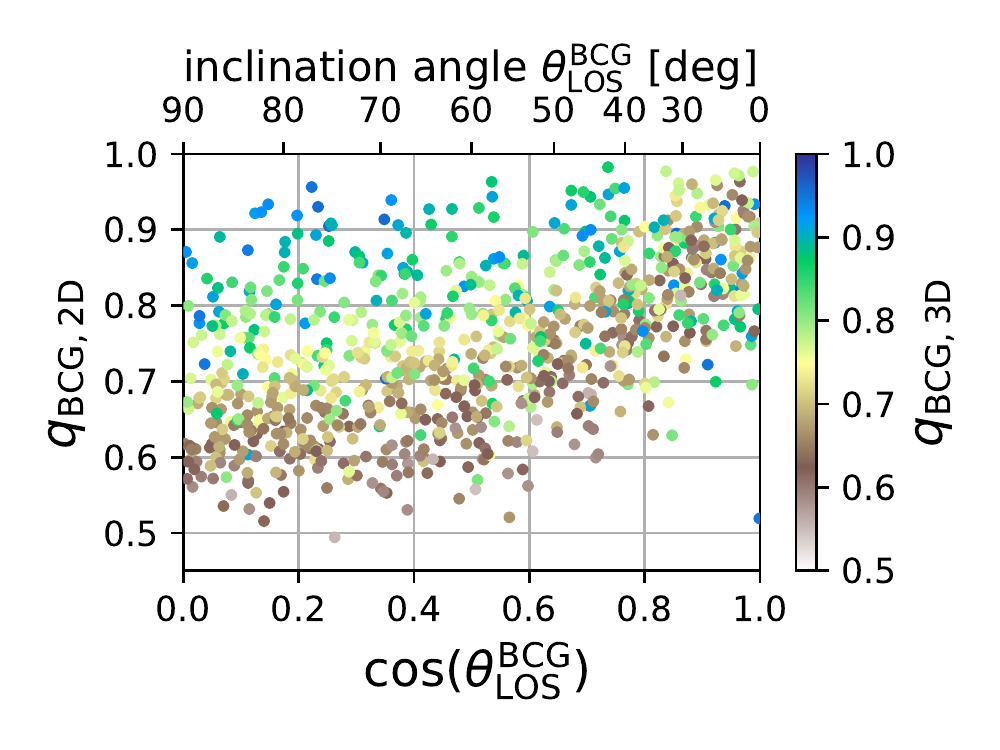}
    \includegraphics[scale=0.85]{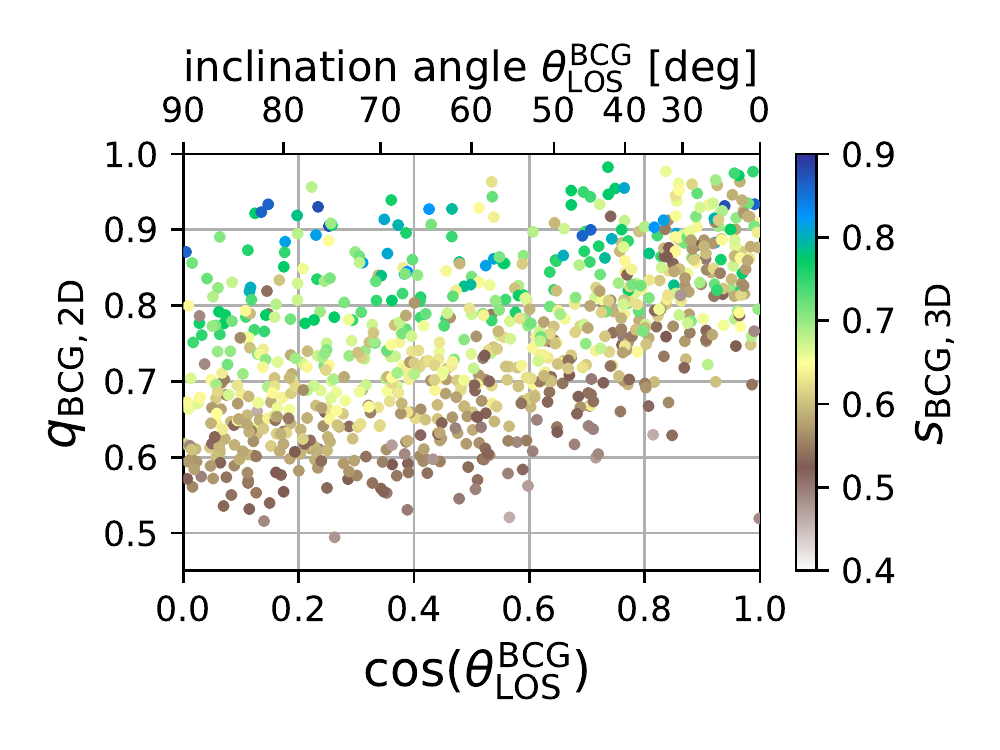}
    \caption{Both panels show the relation between the inclination angle of the BCG and the axis-ratio of the projected BCG stellar mass distribution. Colours indicate the ratios $q=b/a$ (left panel) and $s=c/a$ (right panel) of the full 3D mass distribution. There is a clear trend between inclination angle and 2D axis-ratio for the elongated $q_\mathrm{3D}\leq0.75$ and $s_\mathrm{3D}\leq0.75$ BCGs, but at larger values of $q_\mathrm{3D}$ there is more scatter. }
    \label{fig:qbcg_inclangle}
\end{figure*}

The inclination angle of the BCG provides a direct link to the weak-lensing mass bias, but it can not be measured by observers and a projected observable is required.
Naturally, the ellipticity and orientation of a distribution of particles in 3D is correlated to the ellipticity projected to 2D. 
For a prolate spheroid $a>b = c$, a projection along the major axis results in a round 2D shape, and a projection along any of the two minor axes would show an elliptical 2D shape, with the ellipticity increasing going from $\theta=0$ to 90 degrees.
In Figure~\ref{fig:qbcg_inclangle} we compare the 3D observables measured within 50 kpc to the 2D axis-ratio $q_{2D}$, the only parameter available for observers, for all unflagged BCGs. \rr{We use the method described in Section~\ref{sec:3dshapes} to calculate the 2D axis-ratio from the three-dimensional data.} In the left panel, the trend we have described is clearly exhibited by the yellow-brown ($q_\mathrm{3D}<0.75$) \rr{data points}. At $b/a > 0.75$ this trend is still there, but there is also more scatter in the \rr{data points}. 
\rr{Quantitatively, the elongated BCGs (those with $q_\mathrm{3D}<0.75$) have more correlation between the observed shape $q_\mathrm{2D}$ and the inclination angle $\theta^\mathrm{BCG}_\mathrm{LOS}$ with a Spearman correlation coefficient of $\sim0.76$, compared to the intrinsically spherical BCGs (those with $q_\mathrm{3D}>0.75$), which have a coefficient of $\sim0.15$. Intuitively, this is to be expected.}
Spherical objects ($a \sim b \sim c$) will always look round in projection regardless of the inclination angle and introduce scatter at $q_\mathrm{2D} > 0.9$. Oblate spheroids ($a \sim b >c$) will look elliptical in 2D when projected along the major axis and remain elliptical when rotating the LOS towards the medium axis $b$, but will appear increasingly rounder when rotating the LOS towards the minor axis $c$. As such, they produce an opposite trend to prolate spheroids. This behaviour is most notable for the blue ($a \sim b$) data point in the bottom right of the left panel at cos($\theta)=1$, which appears to be elliptical in projection. In the right panel this BCG is shown to have a small $c/a$, thus it is an oblate spheroid.

The BCGs will likely not be perfectly spherical, prolate or oblate, but their triaxial mass distributions ($a>b>c$) will tend towards any of these three. The right panel of Figure~\ref{fig:qbcg_inclangle} shows that the simulations contain no spherical BCGs as $s_\mathrm{3D}<0.9$ for all BCGs. 
We find that the large majority of our BCGs are prolate spheroids, 249 out of the 316 BCGs have $b$ closer in value to $c$ than to $a$. This is consistent with the observational study by \citet{fasano10}.

Therefore, when observing an elliptical BCG (in 2D), it is very likely that it is in fact an elongated BCG (in 3D) oriented roughly in the plane of the sky. \ir{Note, if observing a round BCG (in 2D), the BCG may either be truly round (in 3D) or elongated with its semi-major axis aligned with the line of sight.  The latter is more likely, given the predominance of prolate BCGs.}

\subsection{Projected BCG shapes}

In an ideal case, the projected shape of the BCG holds information of the inclination angle. In this subsection, we examine BCG shapes measured from the projected stellar density maps.  These shape measurements are more in line with measurements an observer could make. Since substructures in the stellar density map might alter the measured shape of the BCG, we employ three different shape measurements as a consistency check with methods described in Section~\ref{sec:2dshapes}.

\subsubsection{Shape measurement comparison}
\label{sec:2dshape_comparison}

\begin{figure}
    \centering
    \includegraphics[width=\columnwidth]{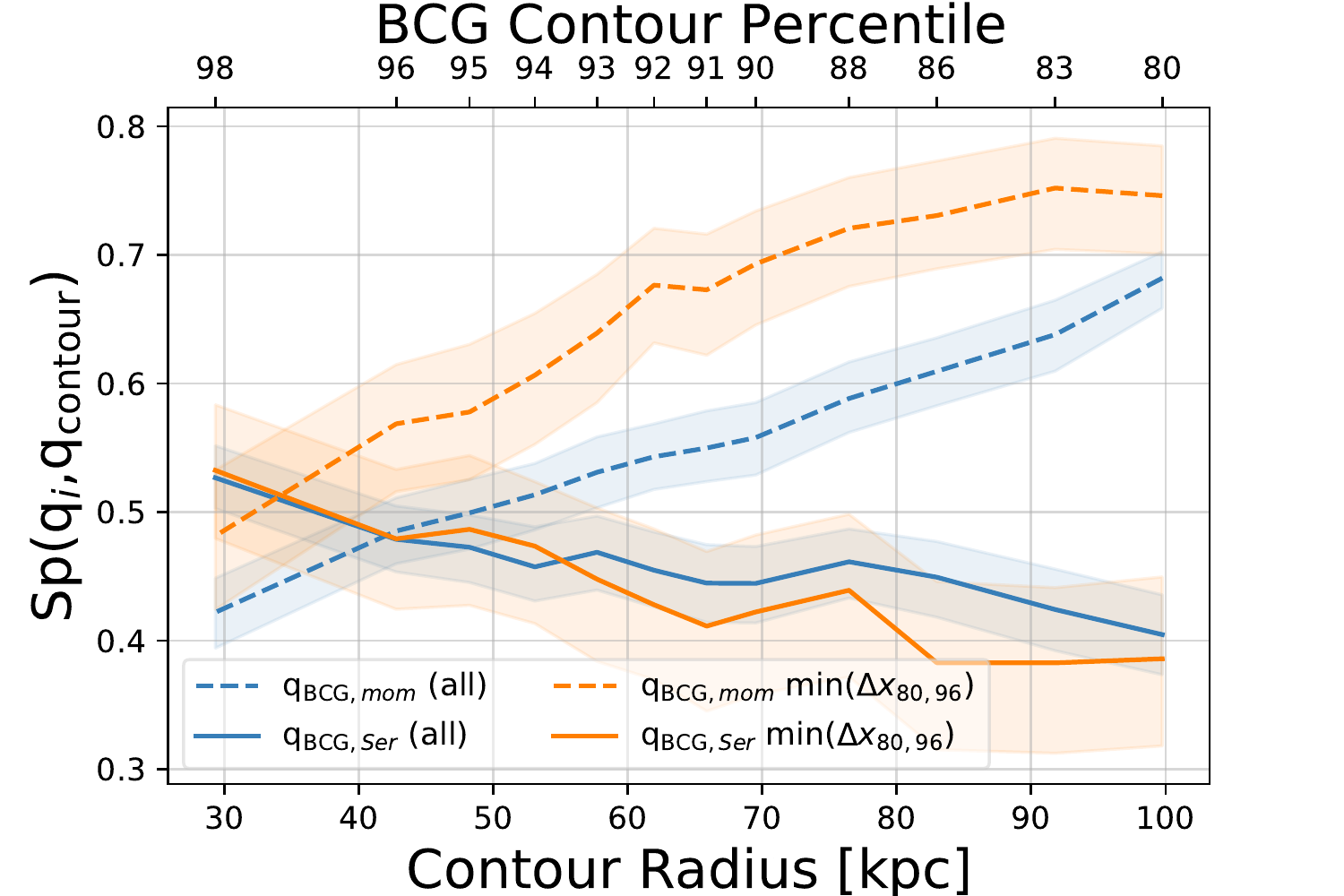}
    \caption{Consistency between 2-dimensional BCG shape measurement methods: Spearman correlation strength between the BCG shape measured with either the Sers\'{i}c profile fit (solid lines) or the unweighted moments (dashed lines) and the 2-dimensional BCG shape measured at different contour percentiles as a function of the median radial size of the shape measurement.  Blue lines indicate measurements for the entire sample, and orange for the subset of images in the lowest quartile of centroid offsets measured between the 80th and 96th percentile contours. Top axis label shows the corresponding brightness percentile defining the contour at which we measure the BCG shape.  Shape measurements that enclose more of the BCG (lower percentile contours) better correlate with shapes measured with the unweighted moment method.  Shape measurements focused on the center of the BCG (higher percentile contours) better correlate with shapes measured with the Sers\'{i}c profile fit.  Note, measurements for all contours at 95th percentile isophote and larger only include BCGs where the contour shape measurement satisfies our quality flag.
    }
    \label{fig:qbcg_qi}
\end{figure}

We show the comparison between our three shape measurement methods in Figure~\ref{fig:qbcg_qi}, which shows the Spearman correlation strength between shapes from either the image moment measurements (dashed lines) or the Sers\'{i}c fits (solid lines) and the shape from a isophote contour, as a function of the isophote contour percentile.  We show the percentile value on the top x-axis labels and the corresponding median radius of that percentile contour on the bottom x-axis labels.  The blue lines show the correlation for the shape measurements when calculated for the galaxies passing the quality cut. For the 96th and 98th percentile isophote contours, where our quality flag $\Delta x_{i,96}$ is undefined, the blue line shows measurements for all BCGs. The shaded region illustrates the 16-84 percentile error. 
The Spearman correlation strength between different BCG shape measurement methods varies between $\sim 0.4$, which is a moderate correlation, and $\sim 0.7$, a strong correlation.  The largest contour shape measurements ($r\approx100$~kpc) strongly correlate with the image moment shape measurements.

In Table~\ref{tab:spcorr}, we summarize the correlation strengths between BCG contour shape measurements methods at percentiles $p=[80,88,96]$.  We show the Spearman correlation for different selections: clusters with unflagged shape measurements, a quarter of the clusters with the lowest normalized centroid offset $\Delta x_{80,96}$, and relaxed clusters according to the \citet{deluca20} and the \citet{cui18} criteria.  

Figure~\ref{fig:qbcg_qi} quantifies the behaviour visible in the two examples of Figure~\ref{fig:bcg_shapes}. Sers\'ic models (solid lines) tend to have small half-light radii because the fit is dominated by the massive BCG core. Sers\'ic derived shape measurements therefore mostly describe the shape of the BCG core. Hence, the correlation between $q_\mathrm{2D,Ser}$ and $q_\mathrm{2D, con}$ (solid blue line) is strongest for the smallest contours. On the other hand, unweighted moments (dashed lines) are sensitive to all mass in the image and mostly to masses at large radii from the centre (see Equation~\ref{eq:moments}). This relationship leads to a similar performance between the image moment shape measurements and the shape measurements from the largest contours.  Both trace the mass far from the centre, and thus the correlation $Sp(q_\mathrm{2D, mom},q_\mathrm{2D,con})$ (dashed blue line) is strong for the biggest contours.


\subsubsection{Shapes compared to observations}


Despite the difficulty of simulating realistic central galaxies, simulations are able to reproduce stellar mass profiles fairly well \citep{ardila21}. However, as a consistency check we compare our measured shapes to values found in observations. The range of axis-ratios in observations is $\sim0.4-$1.0 \citep{fasano10, marrone12, herbonnet19}. We find that our measured BCG shapes are on average more elliptical\ir{, likely due to our choice in shape measurement method}. The unweighted moments have a range of $q_\mathrm{2D}$ similar to observations and a median at $q_\mathrm{2D} \approx 0.65$. Contour measurements result in more elliptical values, from $q_\mathrm{2D}\approx0.3$ to 1.0, with a median at $\approx$0.45. But, we emphasize that Figure~\ref{fig:qbcg_qi} shows that rank ordering is preserved between moments and contours.  The Spearman correlation between the moments and contour shape measurements is strong, showing that both will measure more elliptical distributions as having smaller values of $q_{2D}$ relative to the entire sample.  The exact value of $q_{2D}$ is less relevant to our analysis, as we are most interested in how the shape measurement scales with weak lensing mass.

\subsubsection{An optical relaxation selection\ir{: min$\Delta x_{80,96}$}}

Naively, we would expect different shape measurement methods to be more consistent with one another for the most relaxed clusters, which have a minimum of massive substructures \citep{lauer14,goldenmarx21}.  As a proxy for this in projected stellar density space, we checked whether a stricter cut on the centroid offset would further improve the correlation illustrated in the blue lines in Figure~\ref{fig:qbcg_qi}. We find that keeping only a quarter of the data with the smallest $\Delta x_{80,96}$ consistently led to stronger correlations. These are shown as the orange lines in Figure~\ref{fig:qbcg_qi}. At all radii this cut largely improved the Spearman correlation strength between shapes measured using moments and contours. However, the Sers\'ic measurements do not show the same increase in correlation strength using this stricter selection. The Sers\'ic fits mainly trace the shape of the BCG core and are mostly unaffected by mass far from the BCG centre, to which $\Delta x_{80,96}$ is sensitive.

We motivate the selection of clusters whose centroids are least offset from one another as a proxy to relaxation criteria that are only available in 3-dimensional data of simulations.  Clusters whose inner and outer contour centroids are least offset will tend to be clusters whose outer contour measurement is not disrupted by substructures and whose center of mass roughly sits at the peak of the light distribution.  

Given that the 80th percentile contour most strongly correlates with the image moment measurements and that the 96th percentile contour closely surrounds the peak of the light distribution, we can consider $\min(\Delta x_{80,96})$ to be a light proxy of the relaxation criteria identified using $\delta x_{\mathrm{halo,CoM}}$ and $f_{\mathrm{sub}}$ (see Section~\ref{sec:relax}).  This offset measurement is \ir{somewhat} sensitive to both, given that an abundance of substructures outside the center will impact the shape measurement at larger contours.  Note, our 2-dimensional selection criterion has some overlap with the relaxation criteria in 3-dimensions; since substructures outside of center can shift the centroid of the 80th percentile contour away from the centroid of the 96th percentile contour, the 2-dimensional selection criterion likely identifies some clusters with high values of $\delta x_{\mathrm{halo,CoM}}$ and $f_{\mathrm{sub}}$.  \ir{Specifically, we found that $\min(\Delta x_{80,96})$ exhibited respective Spearman correlation strengths with $f_{\mathrm{sub}}$ and $\delta x_{\mathrm{halo,CoM}}$ of 0.3 and 0.23 respectively.  We note that the 3-dimensional relaxation criteria are measured at much larger radii than our contours, so are not an exact proxy of one another.  In fact, the correlation strength becomes negligible for 3-dimensional relaxation parameters measured at $r_{200c}$.  But, many of the objects selected by our 2-dimensional criterion} are objects also excluded by the \citet{deluca20} and \citet{cui18} criteria; \ir{respectively, 74\% and 86\% of the objects excluded by our 2-dimensional criterion would have been excluded by the \citet{deluca20} or \citet{cui18} selection.} \rr{Our 2-dimensional criterion identifies $\sim$40\% of the relaxed clusters identified by either \citet{deluca20} and \citet{cui18}.}  We also note that correlations between the BCG shape and the weak lensing mass bias also strengthens when we subselect clusters whose centroids of the 80 and 96th percentiles are least offset.  We further discuss the relationship between BCG shape and the weak lensing mass bias in Section~\ref{sec:bcgwlcorr}.

\subsubsection{Relation between BCG shape and halo shape}

We expect most of the BCG accreted stars to be deposited in the BCG outskirts \citep[e.g.][]{oogi13}.  Hence, the outskirts should be more informative of the clusters assembly history. Indeed, we find that the axis-ratio of larger contours correlate more strongly with the shape of the projected mass density of the whole halo on scales of $\sim R_{200c}$. We show the Spearman correlation strength as a function of BCG radius in Table~\ref{tab:spcorr}. In all four cluster samples, the correlation between the BCG and the halo grows stronger with increasing radius for both the axis ratio and the orientation.  We therefore \ir{conclude} that moments and large contours provide the most useful estimates of the BCG+ICL shape in tracing the underlying halo distribution. 

\subsection{Correlation between BCG shape and weak-lensing mass}\label{sec:bcgwlcorr}

\begin{table*}

    \centering
    \begin{tabular}{c|c|c|c|c|c|c|c|c}
    
    selection & $p$ & $\langle r \rangle$ [kpc] & $Sp(q^\mathrm{con}_\mathrm{BCG},q^\mathrm{mom}_\mathrm{BCG})$ & $Sp(q^\mathrm{con}_\mathrm{BCG},q^\mathrm{Ser}_\mathrm{BCG})$ & $Sp(q_\mathrm{BCG},M^\mathrm{WL}/M)$ & $Sp(q_\mathrm{BCG},q_\mathrm{halo})$ & $Sp(o_\mathrm{BCG},o_\mathrm{halo})$ & clusters \\
\hline
      &  95  &  $48.2^{+4.0}_{-2.0}$ & $0.50 \pm 0.03$  & $0.47 \pm 0.03$  & $0.10 \pm 0.03$  & $0.38 \pm 0.03$  & $0.13 \pm 0.03$  & 935 \\
 good &  88  &  $76.2^{+3.8}_{-2.3}$ & $0.59 \pm 0.03$  & $0.46 \pm 0.03$  & $0.14 \pm 0.04$  & $0.40 \pm 0.04$  & $0.20 \pm 0.03$  & 798 \\
      &  80  &  $99.2^{+3.4}_{-3.1}$ & $0.68 \pm 0.02$  & $0.41 \pm 0.03$  & $0.17 \pm 0.03$  & $0.42 \pm 0.04$  & $0.27 \pm 0.03$  & 757 \\
\hline
      &  95  &  $49.0^{+1.8}_{-1.3}$ & $0.57 \pm 0.05$  & $0.45 \pm 0.06$  & $0.21 \pm 0.06$  & $0.43 \pm 0.07$  & $0.24 \pm 0.06$  & 243 \\
 \ir{(min($\Delta x_{80,96}$)} &  88  &  $76.9^{+2.6}_{-1.4}$ & $0.68 \pm 0.04$  & $0.37 \pm 0.06$  & $0.24 \pm 0.06$  & $0.45 \pm 0.07$  & $0.29 \pm 0.06$  & 243 \\
      &  80  &  $99.0^{+3.0}_{-2.5}$ & $0.71 \pm 0.04$  & $0.35 \pm 0.06$  & $0.29 \pm 0.06$  & $0.45 \pm 0.07$  & $0.26 \pm 0.06$  & 243 \\
\hline
   &  95  &  $48.6^{+3.1}_{-1.8}$ & $0.49 \pm 0.05$  & $0.45 \pm 0.06$  & $0.23 \pm 0.06$  & $0.47 \pm 0.06$  & $0.20 \pm 0.06$  & 271 \\
 relaxed DL21&  88  &  $76.6^{+2.3}_{-1.9}$ & $0.58 \pm 0.05$  & $0.45 \pm 0.06$  & $0.32 \pm 0.06$  & $0.56 \pm 0.06$  & $0.26 \pm 0.06$  & 230 \\  
 \citet{deluca20}  &  80  &  $99.4^{+2.8}_{-2.6}$ & $0.67 \pm 0.04$  & $0.39 \pm 0.06$  & $0.32 \pm 0.06$  & $0.54 \pm 0.07$  & $0.30 \pm 0.06$  & 225 \\
\hline
  &  95  &  $48.8^{+2.2}_{-1.8}$ & $0.42 \pm 0.07$  & $0.52 \pm 0.07$  & $0.26 \pm 0.07$  & $0.58 \pm 0.08$  & $0.16 \pm 0.08$  & 144 \\
 relaxed C18 &  88  &  $76.6^{+2.3}_{-1.8}$ & $0.51 \pm 0.07$  & $0.49 \pm 0.07$  & $0.35 \pm 0.08$  & $0.63 \pm 0.08$  & $0.31 \pm 0.07$  & 132 \\
    \citet{cui18} &  80  &  $99.6^{+3.1}_{-2.9}$ & $0.63 \pm 0.06$  & $0.37 \pm 0.08$  & $0.33 \pm 0.08$  & $0.66 \pm 0.08$  & $0.34 \pm 0.09$  & 130 \\

    \end{tabular}
    \caption{Table of Spearman correlations ($Sp$) between measured projected quantities at various percentile isophotes ($p$), which correspond to different radii from the BCG centre ($r$). This radius was obtained as the average over all clusters in the selection. $q_\mathrm{BCG}$ is the axis-ratio of the contour at percentile $p$, and $q^\mathrm{mom}_\mathrm{BCG}$ and  $q^\mathrm{Ser}_\mathrm{BCG}$ are the axis-ratios measured using the moments and Sers\'ic fits, respectively. The subscript $_\mathrm{halo}$ indicates contour measurements of the halo at $R_{200c}$ and $o$ denotes the projected orientation. The number of clusters in each selection at each percentile isophote are shown in the last column. }
    \label{tab:spcorr}
\end{table*}

\begin{figure}
    \centering
    \includegraphics[width=\columnwidth]{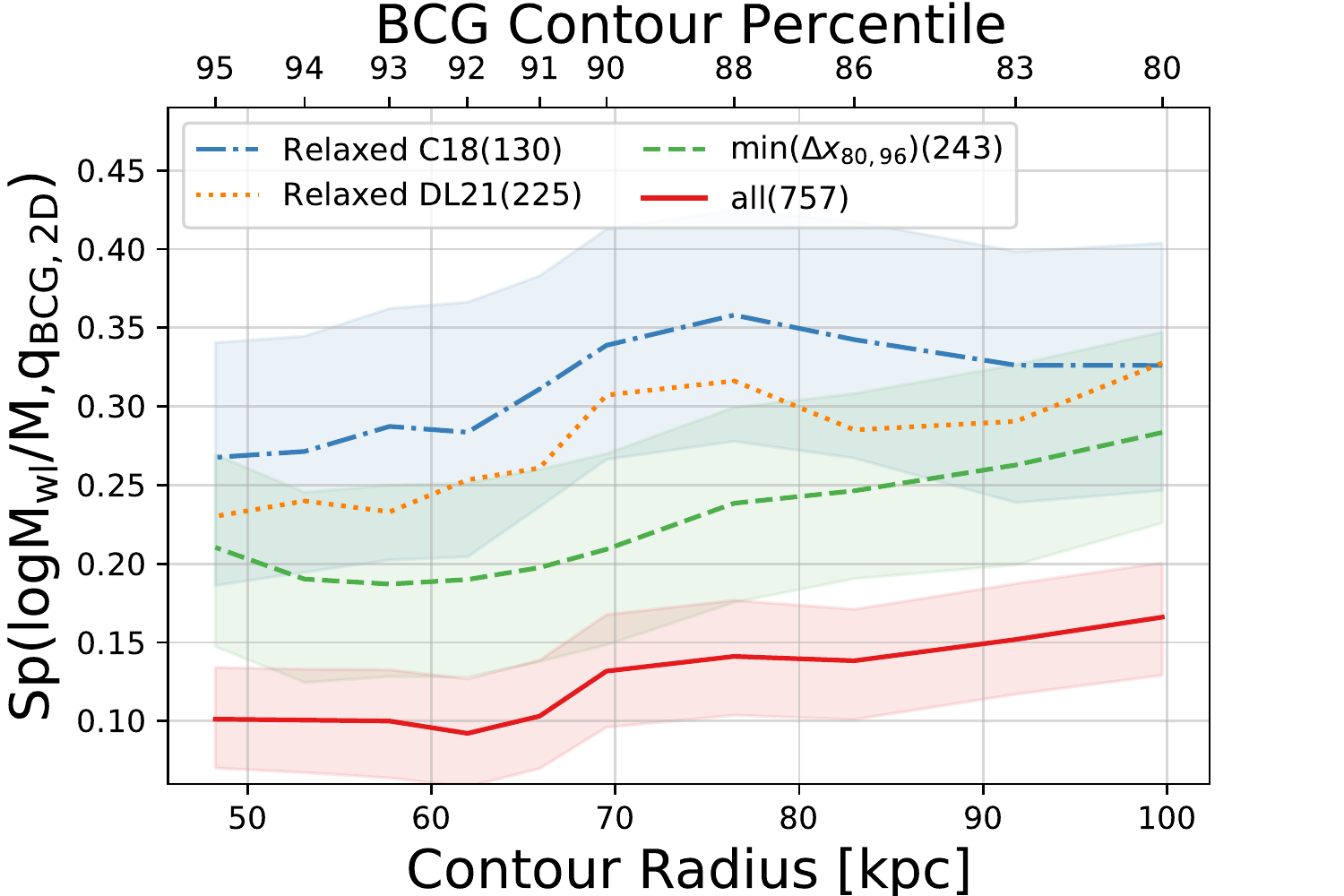}
    \caption{Spearman correlation strength between the weak lensing mass bias and the 2-dimensional BCG shape measured at different contour percentiles as a function of the median radial size of the shape measurement.  Top axis label shows the corresponding brightness percentile defining the contour at which we measure the BCG shape.  Shape measurements that enclose more of the BCG (lower percentile contours) better correlate with the weak lensing mass bias.  Each line corresponds to a different measure of ``relaxedness'' including a selection of the clusters whose 80th and 96th percentile contour centroids are the least offset.  Relaxed cluster subsamples have stronger correlations between their weak lensing mass bias and 2-dimensional BCG shape.
    }
    \label{fig:Sp_qbcg_MWL_radius}
\end{figure}

In Figure~\ref{fig:Sp_qbcg_MWL_radius} we show the Spearman correlation strength between the weak-lensing mass bias and the axis-ratio measured from the contours at different isophotes. The red solid line shows the results for all cluster projections (which pass the shape quality cut) and shows a very weak correlation between mass bias and BCG shape. The quartile of the cluster projections with the smallest offset between the centroids of the 80th and 96th percentile isophotes, min($\Delta x_{80,96}$) (green dashed line) has a more moderate correlation at all radii, exhibiting a slight upward radial trend. The relaxed clusters according to the \citet{deluca20} (orange dotted line) and the \citet{cui18} criteria (blue dash-dotted line) show the strongest correlation between BCG shape and weak-lensing mass bias. All selections show a stronger correlation with increasing radius, supporting our hypothesis that the BCG envelope is a better indicator for the BCG orientation, and thus the halo orientation, to which the weak-lensing signal is sensitive.

We tested how various choices in centroid offset criteria impacted the correlation between the BCG shape measurement and the weak lensing mass bias.  We found that the choice of $\Delta x_{80,96}$ selects clusters whose outer BCG shape best correlate with the weak lensing mass bias.  The effect is analogous to the selection using 3-dimensional relaxation criteria, but is not as stringent.

\begin{figure}
    \centering
    \includegraphics[scale=0.55]{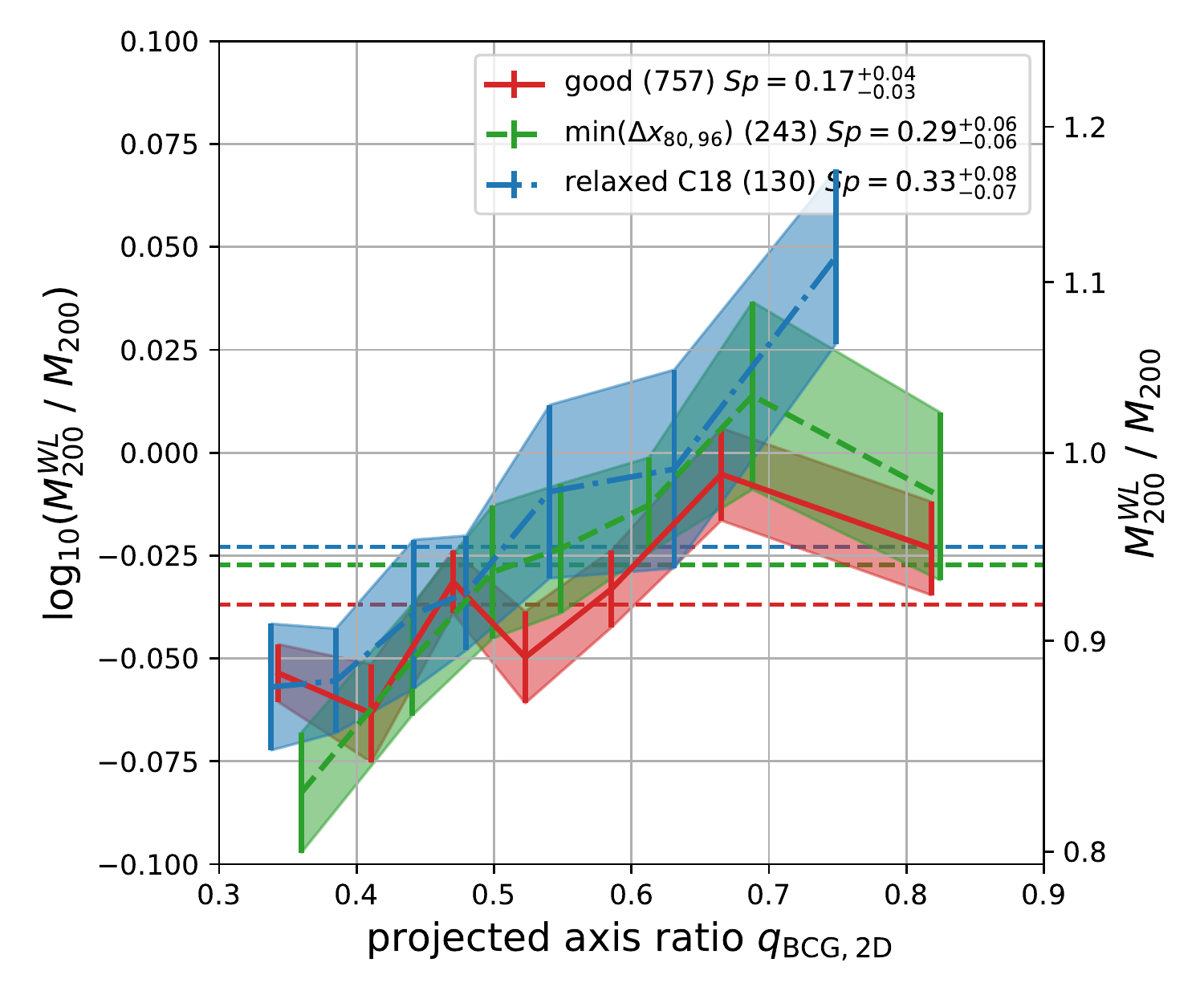}
    \caption{Mean weak lensing mass bias as a function of the axis-ratio of the 80th percentile isophote contour of the projected BCG mass distribution. Lines show the mean mass bias in 7 bins of $q_\mathrm{2D, BCG}$, and the errorbars and the shaded regions show the uncertainty on the mean mass bias. The colours denote the same samples as in Figure~\ref{fig:Sp_qbcg_MWL_radius}. The horizontal lines shows the mean weak-lensing mass bias for each sample in the same colour. 
    Relaxed clusters show the largest correlation between measured BCG shape and mass bias.}
    \label{fig:qbcg_MWL}
\end{figure}

We show the correlation between the BCG shape around 100 kpc (at the 80th percentile isophote) and the weak lensing mass bias in Figure~\ref{fig:qbcg_MWL}. The data was binned into 8 axis-ratio bins with approximately equal numbers of clusters. The coloured lines show for three of the selections in Figure~\ref{fig:Sp_qbcg_MWL_radius} the mean weak-lensing mass bias in bins of the BCG axis-ratio, and the uncertainty on the mean. The horizontal dashed lines show in the same colour scheme for each selection the average weak lensing mass bias. For the full sample (red) the average bias $\langle M^\mathrm{WL}_{200} / M_{200} \rangle \approx -3.5$\% and $\approx -2.5$\% for the other selections.

Compared to the average weak lensing mass bias, there is a clear trend for each selection that the bias is underestimated for elliptical BCGs and overestimated for round BCGs. However, the Spearman correlation strength (\rr{$Sp=0.17\pm0.03$}, shown in the legend) is weak for the full sample in red. The sample of relaxed clusters are shown as the blue dash-dotted line and have a larger, but still only moderate correlation of $Sp=0.33$. The relaxed clusters according to \citet{deluca20} show a very similar behaviour to the blue dash-dotted line. The selection based on $\Delta x_{80,96}$ in green performs similar to the relaxation criteria. \rr{For the highest axis ratio bin, the green and red lines show a slight decrease in mass bias compared to the overall upward trend. This dip is due to spherical BCGs with large values of $q_\mathrm{BCG,2D}$ whose weak-lensing mass estimates are not as high due to cluster orientation (see also Section~\ref{sec:inclination_q}). }

\rr{The scatter about the mean relation of the whole measured sample shown in the red solid line in Figure~\ref{fig:qbcg_MWL} is 19\%. BCG shape information alone provides a moderately tighter scatter compared with the 23\% scatter of weak lensing masses about their true mass.  For relaxed clusters, the scatter about the relation shown as the blue dash-dotted line in Figure~\ref{fig:qbcg_MWL} is 15\%, also a moderately tighter scatter compared with the 20\% of weak lensing masses about their true mass. }

These results imply that observations of relaxed samples of clusters might exhibit the trend between BCG shape and weak-lensing mass bias. 
Therefore, measurements of the BCG shape could improve mass constraints of relaxed clusters. 
BCG shape measurements may not be as indicative of the weak-lensing mass bias in samples that include a larger number of disturbed systems.  Optical proxies of relaxation, similar to the simple $\Delta x_{80,96}$ used here, could be useful in identifying subsamples where the BCG shape and weak-lensing mass bias trend is stronger.  Finally, the relationship between observed outer BCG shape and the halo orientation suggests an indicator for quantifying selection bias in observations.  For example, if the majority of observed BCGs in a sample are fairly round, the sample may be biased with a preferential selection for clusters that are oriented along the line of sight.  In this case, we expect any weak-lensing masses for this sample to bias high compared with the true masses.

\section{Projected BCG mass distribution}
\label{sec:massdistr}

Another possible indicator of BCG orientation is the concentration of mass (or light for observers) in the core compared to the outskirts \citep{giocoli14}. When a mass distribution is viewed along its major axis, it will have more mass projected along the line-of-sight than when viewed from other angles. When viewed along the minor axis, the least amount of mass is projected into the centre of the observed distribution. Here we investigate if the concentration of mass, i.e. the projected mass in the core of BCGs compared with the total stellar mass, correlates with weak lensing mass bias.  Since there is no definition of the total extent of BCGs we instead use the mass within $100$ kpc as an estimate of the total mass. \citet{huang18} show that the mass within 100 kpc is a decent estimate of the total mass of a central galaxy. We look at both our 3D stellar particle data and our 2D projected stellar mass maps to compute the projected stellar mass.

\subsection{Projected stellar mass concentration}
\label{sec:Mconc_theta}

In this section, we examine the relationship between the projected stellar mass concentration and BCG orientation.  To calculate the projected stellar mass concentration, we project our spherically selected BCG stellar particles along each of the three main axes of the simulation box. For each projection, we then select circular apertures of fixed physical radii and sum the mass of all BCG particles within the cylinders. No weights are applied to the particles.

Note, the described procedure likely underestimates projection effects.  We project the $100$~kpc sphere containing BCG particles, and exclude stellar particles outside of the $r=100$~kpc radius.  Hence, projected annuli corresponding to the outer regions of our BCG contain fewer particles than if the projection were of fixed depth at all apertures.  Additionally, we do not tailor the circular apertures to individual BCG mass distributions, which are not always round in projection. We therefore expect projection effects to be underestimated and thus we only use the spherically selected BCG stellar particle data to highlight the connection between the projected stellar mass distribution and the orientation of the BCG. 

\begin{figure*}
    \centering
    \includegraphics[scale=0.85]{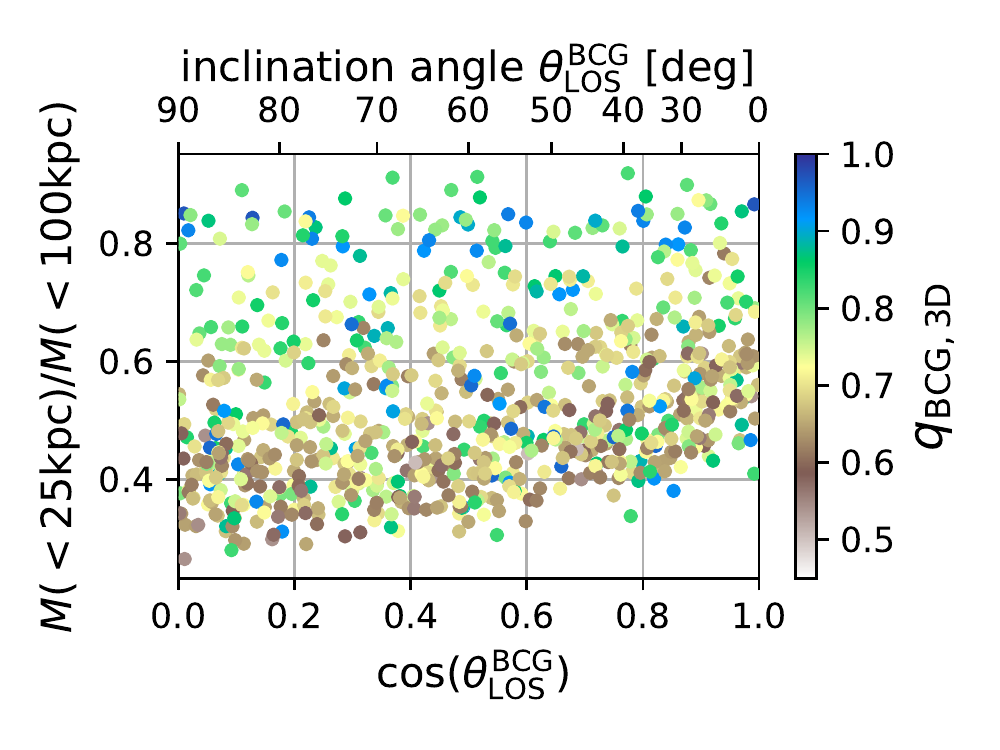}
    \includegraphics[scale=0.85]{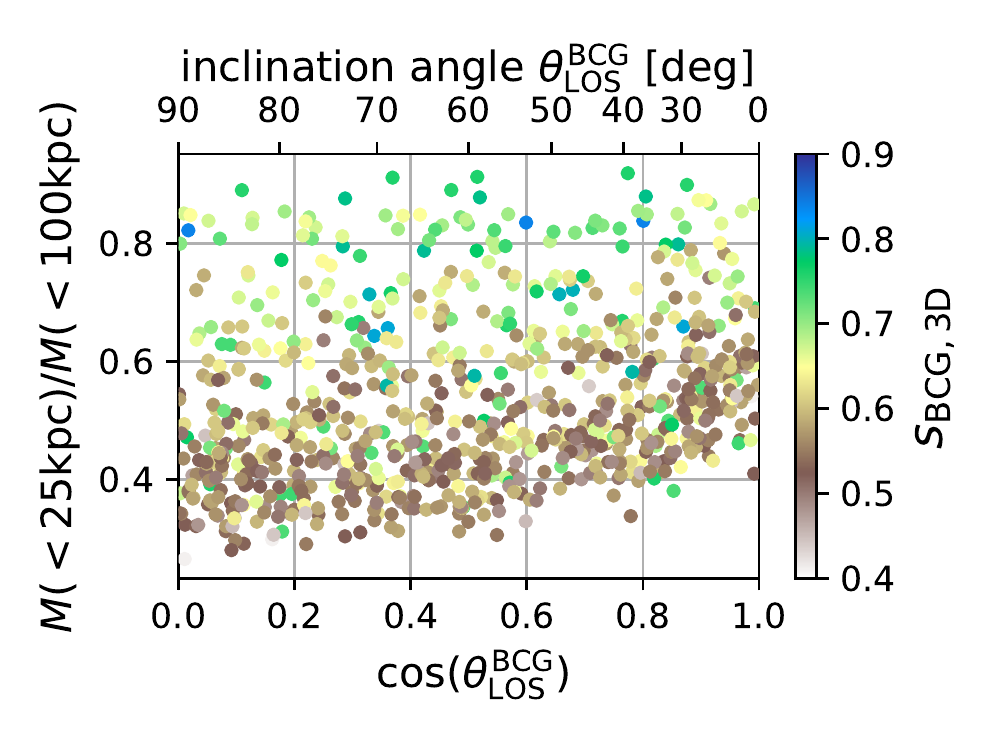}
    \caption{Both panels show the relation between the inclination angle of the BCG and the ratio of projected stellar mass in a circular apertures of radii 25 kpc and 100 kpc. Colours indicate the ratio $q=b/a$ (left panel) and $s=c/a$ (right panel) of the full 3D mass distribution. There is a clear trend between inclination angle and mass concentration for the most elongated $q_\mathrm{3D}\leq0.7$ and $s_\mathrm{3D}\leq0.7$ BCGs, but at larger values of $q_\mathrm{3D}$ there is more scatter.}
    \label{fig:Mconc_theta}
\end{figure*}

We select the 867 BCGs whose 3D shape and orientation measurements were not flagged as contaminated (see Section~\ref{sec:3dshapes}).  Figure~\ref{fig:Mconc_theta} shows the relation between the concentration of stellar mass of the BCG and the inclination angle of the BCG $\theta^\mathrm{BCG}_\mathrm{LOS}$.  We define projected stellar mass concentration as the projected stellar mass within a circular aperture of 25~kpc divided by the projected stellar mass within 100~kpc. The latter is, by construction, all spherically selected BCG stellar particles for each cluster. The inclination angle is calculated using 3D image moments based on all particles within a 100 kpc radius.

As expected, less spherical BCGs (lower values of $q_\mathrm{BCG,3D}$ or $s_\mathrm{BCG,3D}$, shown in brown) show the most significant trend between concentration and inclination angle. BCGs with $q_\mathrm{BCG,3D}<0.7$ have a stellar mass concentration of $\sim0.4$ at cos($\theta^\mathrm{BCG}_\mathrm{LOS}$)$=0$ which rises to almost $0.6$ at  cos($\theta^\mathrm{BCG}_\mathrm{LOS}$)$=1$. There is more mass in the inner 25 kpc when an elongated BCG is viewed along its major axis than when it is viewed along its minor axis.  However, \rr{for} more spherical BCGs (i.e. $q_\mathrm{BCG,3D}>0.7$ and $s_\mathrm{BCG,3D}>0.6$), the relation between concentration and inclination angle exhibits a wider scatter.  For these objects the difference between $a$ and $b, c$ is too small to lead to an observable effect.

In conclusion, the BCGs with the lowest observed concentration, so with relatively more mass in their outskirts compared to the core, will likely not be oriented along the line of sight. BCGs with most of their mass in the inner 25 kpc are likely fairly spherical and could have any inclination angle. Given the trend between BCG concentration and orientation for more elongated BCG, we may expect to see a relation between mass concentration and weak-lensing mass bias.  We explore this possibility in the following subsection.

\subsection{Correlation concentration and weak-lensing mass}

\begin{figure*}
\centering
 \includegraphics[width=.45\linewidth]{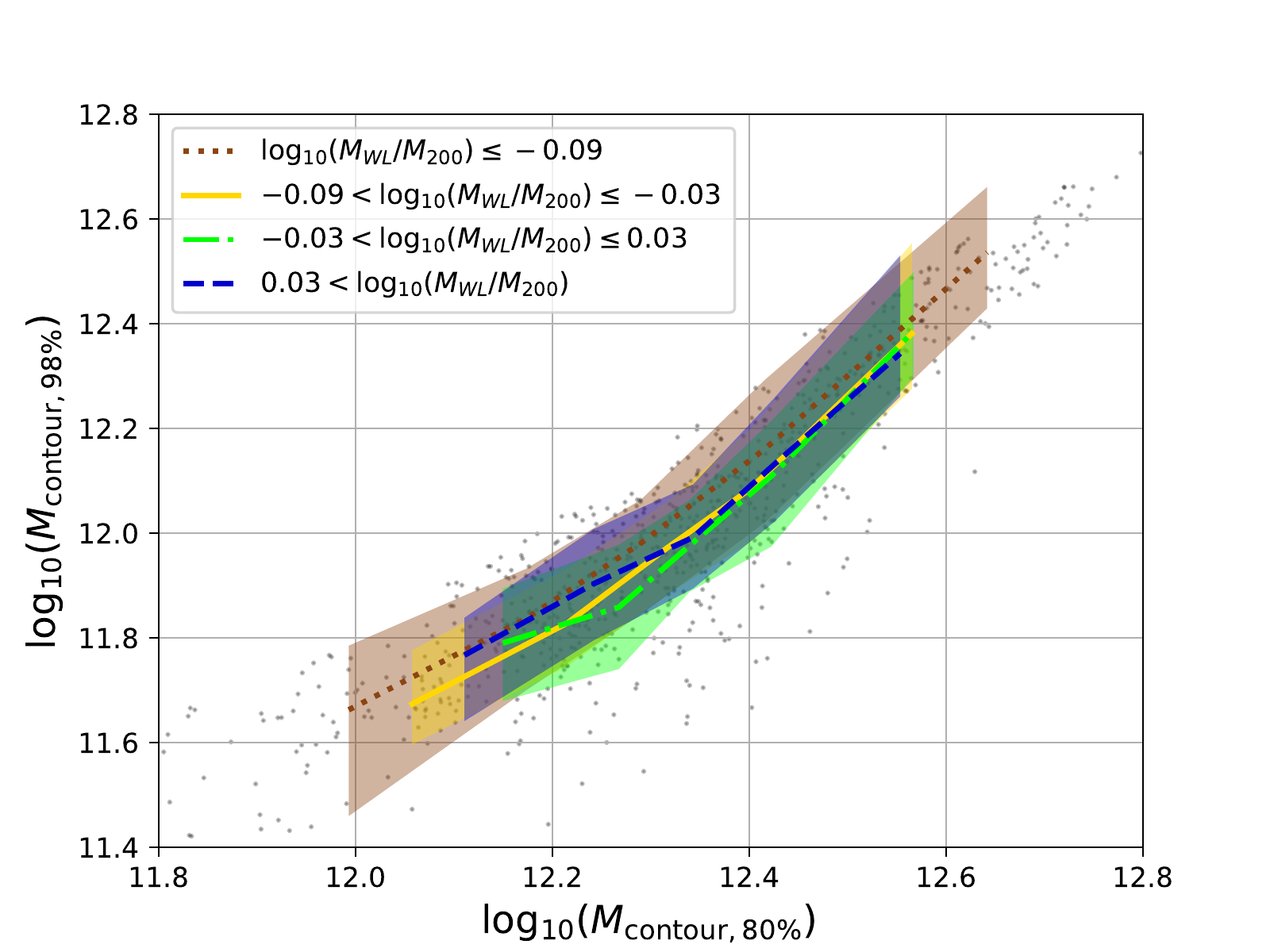}
 \includegraphics[width=.45\linewidth]{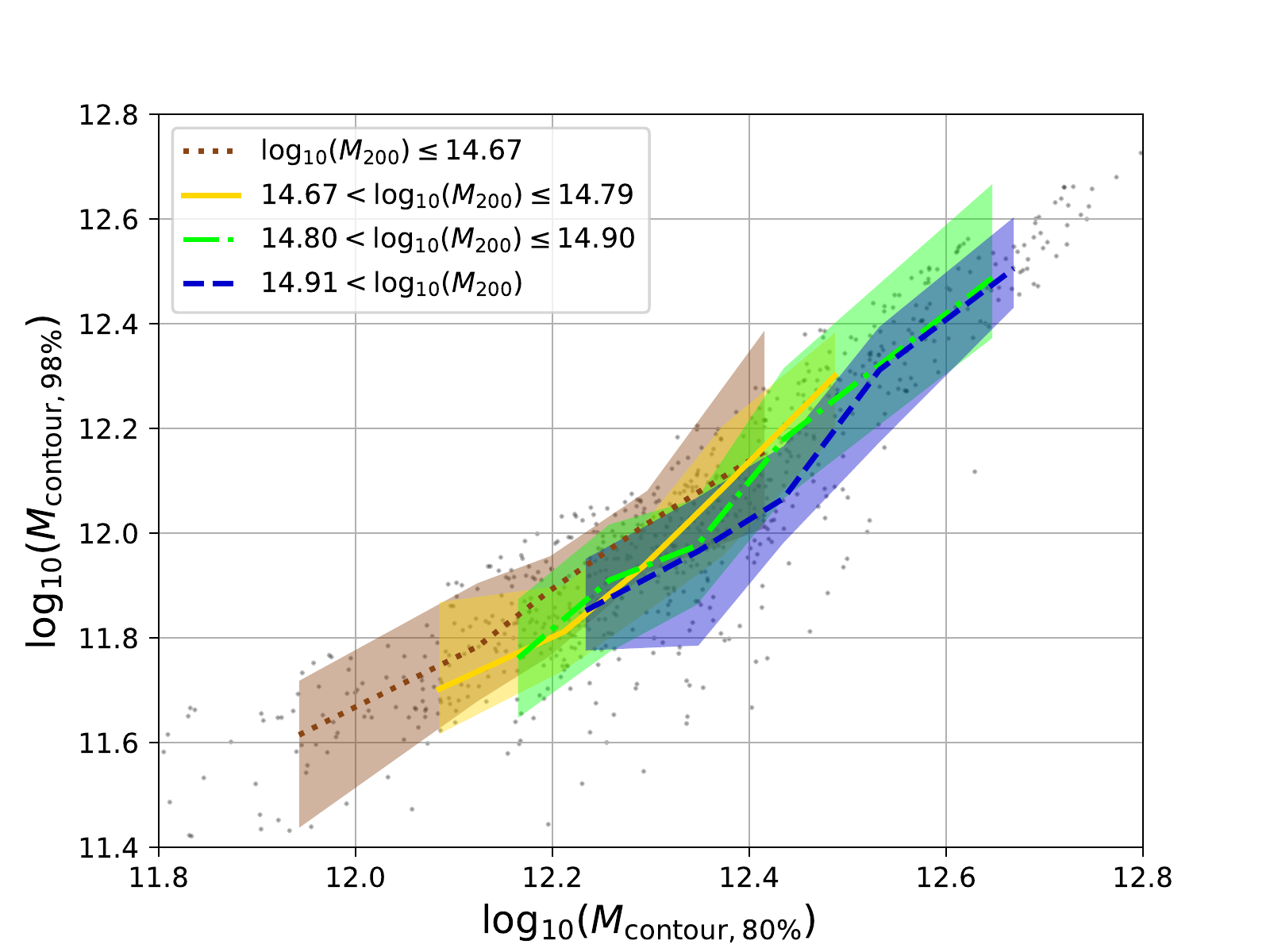}
\caption{Projected stellar mass within the 80\% isophote $M_{\mathrm{contour}, 80\%}$, an estimate of the total stellar mass of the BCG, and the 98\% isophote contours $M_{\mathrm{contour}, 98\%}$, an estimate of the stellar mass of the core of the BCG. Coloured lines show the median behaviour of clusters in four bins in weak-lensing mass bias (\textit{left}) and halo mass (\textit{right}). While there is no visible trend for mass bias there is a trend in halo masses. More massive halos tend to also have less centrally concentrated stellar mass distributions in their central galaxy, as seen in upper-rightward separation of halo mass bins. }
\label{fig:m80-m98-m200-mwl}
\end{figure*}

To quantify the relation between the projected stellar mass concentration and the weak lensing mass bias, we use the 2D projected stellar mass maps. Unlike the 3D particle data, these maps have a uniform depth of 400~kpc over the entire image. We use the contour measurements described in Section~\ref{sec:2dshapes} to capture the projected shape of the galaxy and sum up the mass within a contour to compute the projected stellar mass.

The left panel of Figure~\ref{fig:m80-m98-m200-mwl} shows the mass enclosed in the 80th percentile isophote contour $M_\mathrm{contour, 80\%}$ and the mass enclosed in the 98th percentile isophote contour $M_\mathrm{contour, 98\%}$ for 757 BCGs with $\Delta x_{80,96}<0.2$.
First, we binned the data into 4 bins according to the weak lensing mass bias. Each bin contains a quarter of the total number of projected clusters, approximately 190 projections each. Then, for each quartile of the data we bin the data into 5 bins of $M_\mathrm{contour, 80\%}$ with equal numbers of projections in each bin. We compute the median of $M_\mathrm{contour, 98\%}$ in each bin and show the result as the coloured lines in Figure~\ref{fig:m80-m98-m200-mwl}. The shaded areas show for each line the 16-84 percentile confidence interval in $M_\mathrm{contour, 98\%}$.

In this plane of $M_\mathrm{contour, 80\%}$ and $M_\mathrm{contour, 98\%}$ more concentrated BCGs would lie to the right at fixed $M_\mathrm{contour, 98\%}$, or would lie low at fixed $M_\mathrm{contour, 80\%}$.
Given the relation seen in Section~\ref{sec:Mconc_theta}, we expect clusters with high weak-lensing mass bias to be very concentrated and hence lie on the right side of the distribution. However, there is no distinct behaviour for the four mass bias bins, they are all consistent with each other. Different choices of bins did not change this result, nor did the selection of relaxed haloes with either of the two relaxation criteria.
The absence of a trend is probably due to the large scatter seen in Figure~\ref{fig:Mconc_theta}. The scatter washes out any trend that the most elongated BCGs would show between weak-lensing mass bias and concentration.

In the right panel of Figure~\ref{fig:m80-m98-m200-mwl} we show the same plot, but here the coloured lines represent bins in true halo mass $M_{200c}$. There is a trend that higher mass haloes (e.g. blue dashed line and band) preferentially have less concentrated BCGs (e.g. lower normalization than the brown dotted line and band); more of the mass is in their outskirts. This relation was explored in depth by \citet{huang18b, huang20} in observations and simulations.  Physically, we expect the most massive systems to still be forming.  At these late times, BCGs to accrete mass in their outer envelopes.  The mass distribution of BCGs, quantified by their projected stellar mass concentration, reflects the mass assembly of the cluster.  The consistency between observations and the different simulations supports this physical picture.

Qualitatively the results in Figure~\ref{fig:m80-m98-m200-mwl} did not change with the adoption of a sample of relaxed clusters. We note that our 98\% isophote contour is much larger than the 10 kpc used by \citet{huang20}. We also looked at higher percentile isophote contours, such that we computed the projected stellar mass within a smaller radius, and found very similar results to Figure~\ref{fig:m80-m98-m200-mwl}.

For our sample of simulated clusters the concentration of mass in BCGs is not a good informant on the mass bias in weak-lensing analyses. The stellar mass distribution can nevertheless be a useful tool for cluster studies in optical wavelengths as a proxy for total halo mass.

\section{Conclusions}\label{sec:conclu}

Using the full hydrodynamical resimulated clusters of 'The Three Hundred Project' we studied the mass distributions of galaxy clusters and their central galaxy, also known as the brightest cluster galaxy (BCG). We investigate how the BCG and halo are related and how the BCG can inform weak-lensing studies, which aim to accurately estimate the mass of the halo.

We looked at the alignment between the BCG and the mass distribution of the cluster as a whole. We find that the BCG and the cluster are preferably aligned with on average $\sim$20 degrees between the major axes (Figure~\ref{fig:misalignment_hist}). Relaxed clusters are more tightly aligned for the two relaxation criteria we studied here. Both are based on cluster properties only available to simulators and it remains to be seen how observational relaxation criteria perform.  

The halo-BCG alignment forms the core assumption for the use of the BCG as an indicator of weak-lensing mass bias. In addition, we only employ geometrical arguments to relate observable properties of the BCG to the orientation of the halo. This alignment between central galaxy and host halo has been shown by many different authors for different simulations \citep[e.g.][]{dong14,velliscig15, tenneti15, okabe18, ragone20}. Despite the difficulties in simulating realistic BCGs with properties similar to observations, the fact that different simulations with different physics implementations all show the preferential alignment between the BCG and halo supports the idea that this is a physical phenomenon in galaxy clusters. \rr{We also note that an exercise of masking subhaloes in the simulations would improve the 3-dimensional shape measurements and accomplish stronger correlations the correlation between the BCG shape and weak-lensing mass.}

The triaxial mass distribution of clusters introduces a scatter in the estimated weak lensing mass, which generally assumes a spherical mass distribution \citep[e.g.][]{giocoli12}. The simulated clusters show a direct correlation between weak lensing mass and the orientation of the halo with respect to the line-of-sight. Due to the alignment of BCG and halo, the same correlation is seen for the BCG orientation (Figure~\ref{fig:MWLorientation}). Relaxed clusters have the same Spearman correlation strength whether the BCG or the halo inclination is used as proxy for the mass bias.

We find that most BCGs in the simulation are prolate spheroids. For prolate objects the inclination of the major axis to the observer determines the observed shape projected along the line-of-sight. Hence the BCG shape informs the observer on the BCG orientation, and therefore the orientation bias in the weak-lensing mass. We determined projected BCG shapes and measured the correlation with the weak lensing mass (Figures~\ref{fig:Sp_qbcg_MWL_radius}~and~\ref{fig:qbcg_MWL}). Because BCGs are not perfect prolate spheroids, the correlation is relatively weak. Relaxed clusters show the strongest correlation. This is likely due to the tighter alignment between BCG and halo, as we do not find that BCGs in relaxed clusters are more perfect prolate spheroids than in other clusters.

Observational evidence for correlation between BCG shape and weak lensing-mass has been mostly for X-ray selected cluster samples \citep{mahdavi13, marrone12,herbonnet19}. These samples likely contain more relaxed clusters. A cool core in clusters is a likely indicator of relaxedness and because of their high X-ray luminosity, they are preferentially detected in X-ray observations. Our results are in line with this hypothesis. However, it is encouraging that \citet{gruen14} also found a relation for their 12 clusters selected based on millimeter wavelength observations.
\rr{The correlation signal identified in our work is relatively moderate.  Our work supports the premise that BCG information, such as that indicated in Figure~\ref{fig:qbcg_MWL}, could be used to benefit weak-lensing mass constraints of relaxed clusters for work such as \citet{mantz22}.   }

\rr{Note, we use dark matter criteria to identify relaxed clusters from 3-dimensional criteria.  While there are observational proxies for these parameters, e.g. X-ray peak-BCG position offset as a proxy for $x_\mathrm{off}$ and the magnitude gap as a proxy for $f_\mathrm{sub}$, these do not necessarily have a one-to-one correspondence with the dark matter criteria and are often difficult observational measurements to make.  However, our tests on our 2-dimensional proxy for relaxation based on centroid offsets are potentially applicable to observations, depending on the impact of noise in observed images.  We acknowledge these as additional limitations in linking simulation-based conclusions to what can be extracted from or applied to observations.  However, we note that some observational relaxation criteria, e.g. the SPA criteria in X-ray observations, subselect $\sim10-15\%$ of cluster samples to be relaxed \cite{mantz2015}.}

A second possible observable tracer of BCG orientation is the distribution of stellar mass in the galaxy. There should be a difference in projected mass in the BCG core when projecting along the major axis or the minor axis. We computed the projected concentration of stellar mass as the total mass within a small aperture divided by the total mass of the BCG.
We find that the concentration of mass in BCGs is not a good informant on the mass bias in weak-lensing analyses. Nevertheless, we reproduce the results of \citet{huang20} and found that the BCG mass concentration does correlate well with the true halo mass (Figure~\ref{fig:m80-m98-m200-mwl}). It can therefore still be a valuable tool for weak-lensing studies.

As cluster samples will grow in the coming years with new optical, X-ray and millimeter surveys going online, there is increased pressure to control systematic uncertainties \citep[e.g.][]{sartoris16}. Projection effects can introduce large selection biases in cosmological cluster studies based on weak-lensing mass estimates \rr{\citep{sunayama20,descl20,zhangannis22}}. 
Optical observations have a wealth of information, which is currently not fully utilized by weak-lensing studies, instead relying on assumptions of sphericity for large enough samples of clusters.
Although galaxy cluster physics is not fully understood, there are observables with simple relations to the underlying dark matter halo.
Our work has showed that the BCG can be an indicator for orientation bias in weak-lensing masses. \rr{Alternatively, the distribution of satellite galaxies also traces the halo mass distribution \citep[e.g.][]{velliscig15, ragone20, gonzalez_21,shi21} and might be combined with the BCG shape for a better proxy of cluster orientation. 
} \rr{We leave potential studies to future work.}

Determining a cluster's central galaxies is standard practice and hence almost always available for weak-lensing studies. The BCG therefore provides a cheap way for studies of relaxed clusters to improve their precision and accuracy.  
Unfortunately, the galaxy determined as the central is not always the true central galaxy \citep[e.g.][]{zhang19b}, and this miscentring will wash out the correlation to the halo inclination. The upcoming multiwavelength large-area surveys can provide more than just mass-observable scaling relations, but also accurate cluster centres and reliable central galaxy candidates \citep{george12}. 
A full combination of available data will provide the best way towards to tightest cosmological constraints.

\section*{Acknowledgements}

To the best of our ability we tried to make plots friendly for colourblind readers using the tool from  www.colororacle.org.

This work has been made possible by
the ’The Three Hundred’ collaboration.\footnote{https://www.the300-project.org}. \ir{The simulations used in this paper have been performed in
the MareNostrum Supercomputer at the Barcelona Supercomputing
Center, thanks to CPU time granted by the Red Española de Supercomputación. As part of The Three Hundred Project, this work has received financial support from the European Union’s Horizon 2020 Research and Innovation Programme under the Marie Sklodowska Curie grant agreement number 734374, the LACEGAL project.}

RH and AvdL acknowledge support by the US Department of Energy under award DE-SC0018053.
AC acknowledges support from the National Science Foundation PHY 1852239 ``Summer Undergraduate Research in Physics and Astronomy at the University of Michigan" grant.
CA acknowledges support from the LSA Collegiate Fellowship at the University of Michigan, the Leinweber Foundation, and DoE Award DE-SC009193.
CG and MM acknowledge the grants ASI n.I/023/12/0, ASI-INAF n.  2018-23-HH.0  
and PRIN-MIUR 2017 WSCC32 ``Zooming into dark matter and proto-galaxies with massive lensing clusters; they also acknowledge support from INAF (funding of main-stream projects).  \ir{WC is supported by the European Research Council under grant number 670193 and by the STFC AGP Grant ST/V000594/1. He further acknowledges the science research grants from the China Manned Space Project with NO. CMS-CSST-2021-A01 and CMS-CSST-2021-B01. GY acknowledges financial support from Ministerio de Ciencia, Innovación y Universidades / Fondo Europeo de Desarrollo Regional, under research grant PGC2018-094975-C21.}\\

\section*{Data Availability}

Data available upon request



\bibliographystyle{mnras}
\bibliography{example} 

\begin{thebibliography}{}
\makeatletter
\relax
\def\mn@urlcharsother{\let\do\@makeother \do\$\do\&\do\#\do\^\do\_\do\%\do\~}
\def\mn@doi{\begingroup\mn@urlcharsother \@ifnextchar [ {\mn@doi@}
  {\mn@doi@[]}}
\def\mn@doi@[#1]#2{\def\@tempa{#1}\ifx\@tempa\@empty \href
  {http://dx.doi.org/#2} {doi:#2}\else \href {http://dx.doi.org/#2} {#1}\fi
  \endgroup}
\def\mn@eprint#1#2{\mn@eprint@#1:#2::\@nil}
\def\mn@eprint@arXiv#1{\href {http://arxiv.org/abs/#1} {{\tt arXiv:#1}}}
\def\mn@eprint@dblp#1{\href {http://dblp.uni-trier.de/rec/bibtex/#1.xml}
  {dblp:#1}}
\def\mn@eprint@#1:#2:#3:#4\@nil{\def\@tempa {#1}\def\@tempb {#2}\def\@tempc
  {#3}\ifx \@tempc \@empty \let \@tempc \@tempb \let \@tempb \@tempa \fi \ifx
  \@tempb \@empty \def\@tempb {arXiv}\fi \@ifundefined
  {mn@eprint@\@tempb}{\@tempb:\@tempc}{\expandafter \expandafter \csname
  mn@eprint@\@tempb\endcsname \expandafter{\@tempc}}}

\bibitem[\protect\citeauthoryear{{Abbott} et~al.,}{{Abbott}
  et~al.}{2020}]{descl20}
{Abbott} T.~M.~C.,  et~al., 2020, \mn@doi [\prd] {10.1103/PhysRevD.102.023509},
  \href {https://ui.adsabs.harvard.edu/abs/2020PhRvD.102b3509A} {102, 023509}

\bibitem[\protect\citeauthoryear{{Aguena} et~al.,}{{Aguena}
  et~al.}{2021}]{aguena21}
{Aguena} M.,  et~al., 2021, \mn@doi [\mnras] {10.1093/mnras/stab264}, \href
  {https://ui.adsabs.harvard.edu/abs/2021MNRAS.502.4435A} {502, 4435}

\bibitem[\protect\citeauthoryear{{Allen}, {Evrard}  \& {Mantz}}{{Allen}
  et~al.}{2011}]{allen11}
{Allen} S.~W.,  {Evrard} A.~E.,   {Mantz} A.~B.,  2011, \mn@doi [\araa]
  {10.1146/annurev-astro-081710-102514}, \href
  {https://ui.adsabs.harvard.edu/abs/2011ARA&A..49..409A} {49, 409}

\bibitem[\protect\citeauthoryear{{Ardila} et~al.,}{{Ardila}
  et~al.}{2021}]{ardila21}
{Ardila} F.,  et~al., 2021, \mn@doi [\mnras] {10.1093/mnras/staa3215}, \href
  {https://ui.adsabs.harvard.edu/abs/2021MNRAS.500..432A} {500, 432}

\bibitem[\protect\citeauthoryear{{Baltz}, {Marshall}  \& {Oguri}}{{Baltz}
  et~al.}{2009}]{baltz09}
{Baltz} E.~A.,  {Marshall} P.,   {Oguri} M.,  2009, \mn@doi [\jcap]
  {10.1088/1475-7516/2009/01/015}, \href
  {https://ui.adsabs.harvard.edu/abs/2009JCAP...01..015B} {2009, 015}

\bibitem[\protect\citeauthoryear{{Beck} et~al.,}{{Beck} et~al.}{2016}]{beck16}
{Beck} A.~M.,  et~al., 2016, \mn@doi [\mnras] {10.1093/mnras/stv2443}, \href
  {https://ui.adsabs.harvard.edu/abs/2016MNRAS.455.2110B} {455, 2110}

\bibitem[\protect\citeauthoryear{{Becker} \& {Kravtsov}}{{Becker} \&
  {Kravtsov}}{2011}]{becker11}
{Becker} M.~R.,  {Kravtsov} A.~V.,  2011, \mn@doi [\apj]
  {10.1088/0004-637X/740/1/25}, \href
  {https://ui.adsabs.harvard.edu/abs/2011ApJ...740...25B} {740, 25}

\bibitem[\protect\citeauthoryear{{Bleem} et~al.,}{{Bleem}
  et~al.}{2020}]{bleem20}
{Bleem} L.~E.,  et~al., 2020, \mn@doi [\apjs] {10.3847/1538-4365/ab6993}, \href
  {https://ui.adsabs.harvard.edu/abs/2020ApJS..247...25B} {247, 25}

\bibitem[\protect\citeauthoryear{{Bocquet} et~al.,}{{Bocquet}
  et~al.}{2019}]{bocquet19}
{Bocquet} S.,  et~al., 2019, \mn@doi [\apj] {10.3847/1538-4357/ab1f10}, \href
  {https://ui.adsabs.harvard.edu/abs/2019ApJ...878...55B} {878, 55}

\bibitem[\protect\citeauthoryear{{Boldrin}, {Giocoli}, {Meneghetti}  \&
  {Moscardini}}{{Boldrin} et~al.}{2012}]{boldrin12}
{Boldrin} M.,  {Giocoli} C.,  {Meneghetti} M.,   {Moscardini} L.,  2012,
  \mn@doi [\mnras] {10.1111/j.1365-2966.2012.22120.x}, \href
  {https://ui.adsabs.harvard.edu/abs/2012MNRAS.427.3134B} {427, 3134}

\bibitem[\protect\citeauthoryear{{Boldrin}, {Giocoli}, {Meneghetti},
  {Moscardini}, {Tormen}  \& {Biviano}}{{Boldrin} et~al.}{2016}]{boldrin16}
{Boldrin} M.,  {Giocoli} C.,  {Meneghetti} M.,  {Moscardini} L.,  {Tormen} G.,
   {Biviano} A.,  2016, \mn@doi [\mnras] {10.1093/mnras/stw140}, \href
  {https://ui.adsabs.harvard.edu/abs/2016MNRAS.457.2738B} {457, 2738}

\bibitem[\protect\citeauthoryear{{Ca{\~n}as}, {Lagos}, {Elahi}, {Power},
  {Welker}, {Dubois}  \& {Pichon}}{{Ca{\~n}as} et~al.}{2020}]{canas20}
{Ca{\~n}as} R.,  {Lagos} C. d.~P.,  {Elahi} P.~J.,  {Power} C.,  {Welker} C.,
  {Dubois} Y.,   {Pichon} C.,  2020, \mn@doi [\mnras] {10.1093/mnras/staa1027},
  \href {https://ui.adsabs.harvard.edu/abs/2020MNRAS.494.4314C} {494, 4314}

\bibitem[\protect\citeauthoryear{{Capalbo}, {De Petris}, {De Luca}, {Cui},
  {Yepes}, {Knebe}  \& {Rasia}}{{Capalbo} et~al.}{2020}]{capalbo20}
{Capalbo} V.,  {De Petris} M.,  {De Luca} F.,  {Cui} W.,  {Yepes} G.,  {Knebe}
  A.,   {Rasia} E.,  2020, \mn@doi [\mnras] {10.1093/mnras/staa3900}, \href
  {https://ui.adsabs.harvard.edu/abs/2020MNRAS.tmp.3666C} {}

\bibitem[\protect\citeauthoryear{{Chen}, {Avestruz}, {Kravtsov}, {Lau}  \&
  {Nagai}}{{Chen} et~al.}{2019}]{chen19}
{Chen} H.,  {Avestruz} C.,  {Kravtsov} A.~V.,  {Lau} E.~T.,   {Nagai} D.,
  2019, \mn@doi [\mnras] {10.1093/mnras/stz2776}, \href
  {https://ui.adsabs.harvard.edu/abs/2019MNRAS.490.2380C} {490, 2380}

\bibitem[\protect\citeauthoryear{{Chiu}, {Umetsu}, {Murata}, {Medezinski}  \&
  {Oguri}}{{Chiu} et~al.}{2020}]{chiu20}
{Chiu} I.~N.,  {Umetsu} K.,  {Murata} R.,  {Medezinski} E.,   {Oguri} M.,
  2020, \mn@doi [\mnras] {10.1093/mnras/staa1158}, \href
  {https://ui.adsabs.harvard.edu/abs/2020MNRAS.495..428C} {495, 428}

\bibitem[\protect\citeauthoryear{{Costanzi} et~al.,}{{Costanzi}
  et~al.}{2020}]{costanzi20}
{Costanzi} M.,  et~al., 2020, arXiv e-prints, \href
  {https://ui.adsabs.harvard.edu/abs/2020arXiv201013800C} {p. arXiv:2010.13800}

\bibitem[\protect\citeauthoryear{{Cui} et~al.,}{{Cui} et~al.}{2014}]{cui14}
{Cui} W.,  et~al., 2014, \mn@doi [\mnras] {10.1093/mnras/stt1940}, \href
  {https://ui.adsabs.harvard.edu/abs/2014MNRAS.437..816C} {437, 816}

\bibitem[\protect\citeauthoryear{{Cui}, {Power}, {Borgani}, {Knebe}, {Lewis},
  {Murante}  \& {Poole}}{{Cui} et~al.}{2017}]{cui17}
{Cui} W.,  {Power} C.,  {Borgani} S.,  {Knebe} A.,  {Lewis} G.~F.,  {Murante}
  G.,   {Poole} G.~B.,  2017, \mn@doi [\mnras] {10.1093/mnras/stw2567}, \href
  {https://ui.adsabs.harvard.edu/abs/2017MNRAS.464.2502C} {464, 2502}

\bibitem[\protect\citeauthoryear{{Cui} et~al.,}{{Cui} et~al.}{2018}]{cui18}
{Cui} W.,  et~al., 2018, \mn@doi [\mnras] {10.1093/mnras/sty2111}, \href
  {https://ui.adsabs.harvard.edu/abs/2018MNRAS.480.2898C} {480, 2898}

\bibitem[\protect\citeauthoryear{{Cui} et~al.,}{{Cui} et~al.}{2022}]{cui22}
{Cui} W.,  et~al., 2022, arXiv e-prints, \href
  {https://ui.adsabs.harvard.edu/abs/2022arXiv220214038C} {p. arXiv:2202.14038}

\bibitem[\protect\citeauthoryear{{Curran}}{{Curran}}{2014}]{curran14}
{Curran} P.~A.,  2014, arXiv e-prints, \href
  {https://ui.adsabs.harvard.edu/abs/2014arXiv1411.3816C} {p. arXiv:1411.3816}

\bibitem[\protect\citeauthoryear{{De Luca}, {De Petris}, {Yepes}, {Cui},
  {Knebe}  \& {Rasia}}{{De Luca} et~al.}{2021}]{deluca20}
{De Luca} F.,  {De Petris} M.,  {Yepes} G.,  {Cui} W.,  {Knebe} A.,   {Rasia}
  E.,  2021, \mn@doi [\mnras] {10.1093/mnras/stab1073}, \href
  {https://ui.adsabs.harvard.edu/abs/2021MNRAS.504.5383D} {504, 5383}

\bibitem[\protect\citeauthoryear{{De Lucia} \& {Blaizot}}{{De Lucia} \&
  {Blaizot}}{2007}]{delucia07}
{De Lucia} G.,  {Blaizot} J.,  2007, \mn@doi [\mnras]
  {10.1111/j.1365-2966.2006.11287.x}, \href
  {https://ui.adsabs.harvard.edu/abs/2007MNRAS.375....2D} {375, 2}

\bibitem[\protect\citeauthoryear{{De Propris} et~al.,}{{De Propris}
  et~al.}{2021}]{depropris21}
{De Propris} R.,  et~al., 2021, \mn@doi [\mnras] {10.1093/mnras/staa3286},
  \href {https://ui.adsabs.harvard.edu/abs/2021MNRAS.500..310D} {500, 310}

\bibitem[\protect\citeauthoryear{{Despali}, {Giocoli}  \& {Tormen}}{{Despali}
  et~al.}{2014}]{despali14}
{Despali} G.,  {Giocoli} C.,   {Tormen} G.,  2014, \mn@doi [\mnras]
  {10.1093/mnras/stu1393}, \href
  {https://ui.adsabs.harvard.edu/abs/2014MNRAS.443.3208D} {443, 3208}

\bibitem[\protect\citeauthoryear{{Despali}, {Giocoli}, {Angulo}, {Tormen},
  {Sheth}, {Baso}  \& {Moscardini}}{{Despali} et~al.}{2016}]{despali16}
{Despali} G.,  {Giocoli} C.,  {Angulo} R.~E.,  {Tormen} G.,  {Sheth} R.~K.,
  {Baso} G.,   {Moscardini} L.,  2016, \mn@doi [\mnras]
  {10.1093/mnras/stv2842}, \href
  {https://ui.adsabs.harvard.edu/abs/2016MNRAS.456.2486D} {456, 2486}

\bibitem[\protect\citeauthoryear{{Dietrich} et~al.,}{{Dietrich}
  et~al.}{2014}]{dietrich14}
{Dietrich} J.~P.,  et~al., 2014, \mn@doi [\mnras] {10.1093/mnras/stu1282},
  \href {https://ui.adsabs.harvard.edu/abs/2014MNRAS.443.1713D} {443, 1713}

\bibitem[\protect\citeauthoryear{{Dodelson}, {Heitmann}, {Hirata}, {Honscheid},
  {Roodman}, {Seljak}, {Slosar}  \& {Trodden}}{{Dodelson}
  et~al.}{2016}]{dodelson16}
{Dodelson} S.,  {Heitmann} K.,  {Hirata} C.,  {Honscheid} K.,  {Roodman} A.,
  {Seljak} U.,  {Slosar} A.,   {Trodden} M.,  2016, arXiv e-prints, \href
  {https://ui.adsabs.harvard.edu/abs/2016arXiv160407626D} {p. arXiv:1604.07626}

\bibitem[\protect\citeauthoryear{{Donahue} et~al.,}{{Donahue}
  et~al.}{2016}]{donahue16}
{Donahue} M.,  et~al., 2016, \mn@doi [\apj] {10.3847/0004-637X/819/1/36}, \href
  {https://ui.adsabs.harvard.edu/abs/2016ApJ...819...36D} {819, 36}

\bibitem[\protect\citeauthoryear{{Dong}, {Lin}, {Kang}, {Ocean Wang}, {Dutton}
  \& {Macci{\`o}}}{{Dong} et~al.}{2014}]{dong14}
{Dong} X.~C.,  {Lin} W.~P.,  {Kang} X.,  {Ocean Wang} Y.,  {Dutton} A.~A.,
  {Macci{\`o}} A.~V.,  2014, \mn@doi [\apjl] {10.1088/2041-8205/791/2/L33},
  \href {https://ui.adsabs.harvard.edu/abs/2014ApJ...791L..33D} {791, L33}

\bibitem[\protect\citeauthoryear{{Durret}, {Tarricq}, {M{\'a}rquez}, {Ashkar}
  \& {Adami}}{{Durret} et~al.}{2019}]{durret19}
{Durret} F.,  {Tarricq} Y.,  {M{\'a}rquez} I.,  {Ashkar} H.,   {Adami} C.,
  2019, \mn@doi [\aap] {10.1051/0004-6361/201834374}, \href
  {https://ui.adsabs.harvard.edu/abs/2019A&A...622A..78D} {622, A78}

\bibitem[\protect\citeauthoryear{{Faltenbacher}, {Allgood}, {Gottl{\"o}ber},
  {Yepes}  \& {Hoffman}}{{Faltenbacher} et~al.}{2005}]{faltenbacher05}
{Faltenbacher} A.,  {Allgood} B.,  {Gottl{\"o}ber} S.,  {Yepes} G.,   {Hoffman}
  Y.,  2005, \mn@doi [\mnras] {10.1111/j.1365-2966.2005.09386.x}, \href
  {https://ui.adsabs.harvard.edu/abs/2005MNRAS.362.1099F} {362, 1099}

\bibitem[\protect\citeauthoryear{{Fasano} et~al.,}{{Fasano}
  et~al.}{2010}]{fasano10}
{Fasano} G.,  et~al., 2010, \mn@doi [\mnras]
  {10.1111/j.1365-2966.2010.16361.x}, \href
  {https://ui.adsabs.harvard.edu/abs/2010MNRAS.404.1490F} {404, 1490}

\bibitem[\protect\citeauthoryear{{George} et~al.,}{{George}
  et~al.}{2012}]{george12}
{George} M.~R.,  et~al., 2012, \mn@doi [\apj] {10.1088/0004-637X/757/1/2},
  \href {https://ui.adsabs.harvard.edu/abs/2012ApJ...757....2G} {757, 2}

\bibitem[\protect\citeauthoryear{{Giocoli}, {Meneghetti}, {Ettori}  \&
  {Moscardini}}{{Giocoli} et~al.}{2012}]{giocoli12}
{Giocoli} C.,  {Meneghetti} M.,  {Ettori} S.,   {Moscardini} L.,  2012, \mn@doi
  [\mnras] {10.1111/j.1365-2966.2012.21743.x}, \href
  {https://ui.adsabs.harvard.edu/abs/2012MNRAS.426.1558G} {426, 1558}

\bibitem[\protect\citeauthoryear{{Giocoli}, {Meneghetti}, {Metcalf}, {Ettori}
  \& {Moscardini}}{{Giocoli} et~al.}{2014}]{giocoli14}
{Giocoli} C.,  {Meneghetti} M.,  {Metcalf} R.~B.,  {Ettori} S.,   {Moscardini}
  L.,  2014, \mn@doi [\mnras] {10.1093/mnras/stu303}, \href
  {https://ui.adsabs.harvard.edu/abs/2014MNRAS.440.1899G} {440, 1899}

\bibitem[\protect\citeauthoryear{{Giocoli} et~al.,}{{Giocoli}
  et~al.}{2021}]{giocoli21}
{Giocoli} C.,  et~al., 2021, arXiv e-prints, \href
  {https://ui.adsabs.harvard.edu/abs/2021arXiv210305653G} {p. arXiv:2103.05653}

\bibitem[\protect\citeauthoryear{{Golden-Marx} et~al.,}{{Golden-Marx}
  et~al.}{2021}]{goldenmarx21}
{Golden-Marx} J.~B.,  et~al., 2021, arXiv e-prints, \href
  {https://ui.adsabs.harvard.edu/abs/2021arXiv210702197G} {p. arXiv:2107.02197}

\bibitem[\protect\citeauthoryear{{Gonzalez}, {Ragone-Figueroa}, {Donzelli},
  {Makler}, {Garc{\'\i}a Lambas}  \& {Granato}}{{Gonzalez}
  et~al.}{2021}]{gonzalez_21}
{Gonzalez} E.~J.,  {Ragone-Figueroa} C.,  {Donzelli} C.~J.,  {Makler} M.,
  {Garc{\'\i}a Lambas} D.,   {Granato} G.~L.,  2021, arXiv e-prints, \href
  {https://ui.adsabs.harvard.edu/abs/2021arXiv210703418G} {p. arXiv:2107.03418}

\bibitem[\protect\citeauthoryear{{Gruen} et~al.,}{{Gruen}
  et~al.}{2014}]{gruen14}
{Gruen} D.,  et~al., 2014, \mn@doi [\mnras] {10.1093/mnras/stu949}, \href
  {https://ui.adsabs.harvard.edu/abs/2014MNRAS.442.1507G} {442, 1507}

\bibitem[\protect\citeauthoryear{{Haggar}, {Gray}, {Pearce}, {Knebe}, {Cui},
  {Mostoghiu}  \& {Yepes}}{{Haggar} et~al.}{2020}]{haggar20}
{Haggar} R.,  {Gray} M.~E.,  {Pearce} F.~R.,  {Knebe} A.,  {Cui} W.,
  {Mostoghiu} R.,   {Yepes} G.,  2020, \mn@doi [\mnras]
  {10.1093/mnras/staa273}, \href
  {https://ui.adsabs.harvard.edu/abs/2020MNRAS.492.6074H} {492, 6074}

\bibitem[\protect\citeauthoryear{{Henson}, {Barnes}, {Kay}, {McCarthy}  \&
  {Schaye}}{{Henson} et~al.}{2017}]{henson17}
{Henson} M.~A.,  {Barnes} D.~J.,  {Kay} S.~T.,  {McCarthy} I.~G.,   {Schaye}
  J.,  2017, \mn@doi [\mnras] {10.1093/mnras/stw2899}, \href
  {https://ui.adsabs.harvard.edu/abs/2017MNRAS.465.3361H} {465, 3361}

\bibitem[\protect\citeauthoryear{{Herbonnet}, {von der Linden}, {Allen},
  {Mantz}, {Modumudi}, {Morris}  \& {Kelly}}{{Herbonnet}
  et~al.}{2019}]{herbonnet19}
{Herbonnet} R.,  {von der Linden} A.,  {Allen} S.~W.,  {Mantz} A.~B.,
  {Modumudi} P.,  {Morris} R.~G.,   {Kelly} P.~L.,  2019, \mn@doi [\mnras]
  {10.1093/mnras/stz2913}, \href
  {https://ui.adsabs.harvard.edu/abs/2019MNRAS.490.4889H} {490, 4889}

\bibitem[\protect\citeauthoryear{{Herbonnet} et~al.,}{{Herbonnet}
  et~al.}{2020}]{herbonnet20}
{Herbonnet} R.,  et~al., 2020, \mn@doi [\mnras] {10.1093/mnras/staa2303}, \href
  {https://ui.adsabs.harvard.edu/abs/2020MNRAS.497.4684H} {497, 4684}

\bibitem[\protect\citeauthoryear{{Hilton} et~al.,}{{Hilton}
  et~al.}{2021}]{hilton21}
{Hilton} M.,  et~al., 2021, \mn@doi [\apjs] {10.3847/1538-4365/abd023}, \href
  {https://ui.adsabs.harvard.edu/abs/2021ApJS..253....3H} {253, 3}

\bibitem[\protect\citeauthoryear{{Huang}, {Leauthaud}, {Greene}, {Bundy},
  {Lin}, {Tanaka}, {Miyazaki}  \& {Komiyama}}{{Huang} et~al.}{2018a}]{huang18}
{Huang} S.,  {Leauthaud} A.,  {Greene} J.~E.,  {Bundy} K.,  {Lin} Y.-T.,
  {Tanaka} M.,  {Miyazaki} S.,   {Komiyama} Y.,  2018a, \mn@doi [\mnras]
  {10.1093/mnras/stx3200}, \href
  {https://ui.adsabs.harvard.edu/abs/2018MNRAS.475.3348H} {475, 3348}

\bibitem[\protect\citeauthoryear{{Huang} et~al.,}{{Huang}
  et~al.}{2018b}]{huang18b}
{Huang} S.,  et~al., 2018b, \mn@doi [\mnras] {10.1093/mnras/sty1136}, \href
  {https://ui.adsabs.harvard.edu/abs/2018MNRAS.480..521H} {480, 521}

\bibitem[\protect\citeauthoryear{{Huang} et~al.,}{{Huang}
  et~al.}{2020}]{huang20}
{Huang} S.,  et~al., 2020, \mn@doi [\mnras] {10.1093/mnras/stz3314}, \href
  {https://ui.adsabs.harvard.edu/abs/2020MNRAS.492.3685H} {492, 3685}

\bibitem[\protect\citeauthoryear{{Kasun} \& {Evrard}}{{Kasun} \&
  {Evrard}}{2005}]{kasunandevrard05}
{Kasun} S.~F.,  {Evrard} A.~E.,  2005, \mn@doi [\apj] {10.1086/430811}, \href
  {https://ui.adsabs.harvard.edu/abs/2005ApJ...629..781K} {629, 781}

\bibitem[\protect\citeauthoryear{{Kluge} et~al.,}{{Kluge}
  et~al.}{2020}]{kluge20}
{Kluge} M.,  et~al., 2020, \mn@doi [\apjs] {10.3847/1538-4365/ab733b}, \href
  {https://ui.adsabs.harvard.edu/abs/2020ApJS..247...43K} {247, 43}

\bibitem[\protect\citeauthoryear{{Klypin}, {Yepes}, {Gottl{\"o}ber}, {Prada}
  \& {He{\ss}}}{{Klypin} et~al.}{2016}]{klypin16}
{Klypin} A.,  {Yepes} G.,  {Gottl{\"o}ber} S.,  {Prada} F.,   {He{\ss}} S.,
  2016, \mn@doi [\mnras] {10.1093/mnras/stw248}, \href
  {https://ui.adsabs.harvard.edu/abs/2016MNRAS.457.4340K} {457, 4340}

\bibitem[\protect\citeauthoryear{{Knebe} et~al.,}{{Knebe}
  et~al.}{2020}]{knebe20}
{Knebe} A.,  et~al., 2020, \mn@doi [\mnras] {10.1093/mnras/staa1407}, \href
  {https://ui.adsabs.harvard.edu/abs/2020MNRAS.495.3002K} {495, 3002}

\bibitem[\protect\citeauthoryear{{Knollmann} \& {Knebe}}{{Knollmann} \&
  {Knebe}}{2009}]{knollmannandknebe09}
{Knollmann} S.~R.,  {Knebe} A.,  2009, \mn@doi [\apjs]
  {10.1088/0067-0049/182/2/608}, \href
  {https://ui.adsabs.harvard.edu/abs/2009ApJS..182..608K} {182, 608}

\bibitem[\protect\citeauthoryear{{Lauer}, {Postman}, {Strauss}, {Graves}  \&
  {Chisari}}{{Lauer} et~al.}{2014}]{lauer14}
{Lauer} T.~R.,  {Postman} M.,  {Strauss} M.~A.,  {Graves} G.~J.,   {Chisari}
  N.~E.,  2014, \mn@doi [\apj] {10.1088/0004-637X/797/2/82}, \href
  {https://ui.adsabs.harvard.edu/abs/2014ApJ...797...82L} {797, 82}

\bibitem[\protect\citeauthoryear{{Laureijs} et~al.,}{{Laureijs}
  et~al.}{2011}]{laureijs11}
{Laureijs} R.,  et~al., 2011, arXiv e-prints, \href
  {https://ui.adsabs.harvard.edu/abs/2011arXiv1110.3193L} {p. arXiv:1110.3193}

\bibitem[\protect\citeauthoryear{{Lesci} et~al.,}{{Lesci}
  et~al.}{2020}]{lesci20}
{Lesci} G.~F.,  et~al., 2020, arXiv e-prints, \href
  {https://ui.adsabs.harvard.edu/abs/2020arXiv201212273L} {p. arXiv:2012.12273}

\bibitem[\protect\citeauthoryear{{Li} et~al.,}{{Li} et~al.}{2020}]{li20}
{Li} Q.,  et~al., 2020, \mn@doi [\mnras] {10.1093/mnras/staa1385}, \href
  {https://ui.adsabs.harvard.edu/abs/2020MNRAS.495.2930L} {495, 2930}

\bibitem[\protect\citeauthoryear{{Liu} et~al.,}{{Liu} et~al.}{2021}]{liu21}
{Liu} A.,  et~al., 2021, arXiv e-prints, \href
  {https://ui.adsabs.harvard.edu/abs/2021arXiv210614518L} {p. arXiv:2106.14518}

\bibitem[\protect\citeauthoryear{{Machado Poletti Valle}, {Avestruz}, {Barnes},
  {Farahi}, {Lau}  \& {Nagai}}{{Machado Poletti Valle}
  et~al.}{2020}]{machado20}
{Machado Poletti Valle} L.~F.,  {Avestruz} C.,  {Barnes} D.~J.,  {Farahi} A.,
  {Lau} E.~T.,   {Nagai} D.,  2020, arXiv e-prints, \href
  {https://ui.adsabs.harvard.edu/abs/2020arXiv201112987M} {p. arXiv:2011.12987}

\bibitem[\protect\citeauthoryear{{Mahdavi}, {Hoekstra}, {Babul}, {Bildfell},
  {Jeltema}  \& {Henry}}{{Mahdavi} et~al.}{2013}]{mahdavi13}
{Mahdavi} A.,  {Hoekstra} H.,  {Babul} A.,  {Bildfell} C.,  {Jeltema} T.,
  {Henry} J.~P.,  2013, \mn@doi [\apj] {10.1088/0004-637X/767/2/116}, \href
  {https://ui.adsabs.harvard.edu/abs/2013ApJ...767..116M} {767, 116}

\bibitem[\protect\citeauthoryear{{Mantz} et~al.,}{{Mantz}
  et~al.}{2015a}]{mantz15}
{Mantz} A.~B.,  et~al., 2015a, \mn@doi [\mnras] {10.1093/mnras/stu2096}, \href
  {https://ui.adsabs.harvard.edu/abs/2015MNRAS.446.2205M} {446, 2205}

\bibitem[\protect\citeauthoryear{{Mantz}, {Allen}, {Morris}, {Schmidt}, {von
  der Linden}  \& {Urban}}{{Mantz} et~al.}{2015b}]{mantz2015}
{Mantz} A.~B.,  {Allen} S.~W.,  {Morris} R.~G.,  {Schmidt} R.~W.,  {von der
  Linden} A.,   {Urban} O.,  2015b, \mn@doi [\mnras] {10.1093/mnras/stv219},
  \href {https://ui.adsabs.harvard.edu/abs/2015MNRAS.449..199M} {449, 199}

\bibitem[\protect\citeauthoryear{{Mantz} et~al.,}{{Mantz}
  et~al.}{2022}]{mantz22}
{Mantz} A.~B.,  et~al., 2022, \mn@doi [\mnras] {10.1093/mnras/stab3390}, \href
  {https://ui.adsabs.harvard.edu/abs/2022MNRAS.510..131M} {510, 131}

\bibitem[\protect\citeauthoryear{{Marrone} et~al.,}{{Marrone}
  et~al.}{2012}]{marrone12}
{Marrone} D.~P.,  et~al., 2012, \mn@doi [\apj] {10.1088/0004-637X/754/2/119},
  \href {https://ui.adsabs.harvard.edu/abs/2012ApJ...754..119M} {754, 119}

\bibitem[\protect\citeauthoryear{{Maturi}, {Bellagamba}, {Radovich},
  {Roncarelli}, {Sereno}, {Moscardini}, {Bardelli}  \& {Puddu}}{{Maturi}
  et~al.}{2019}]{maturi19}
{Maturi} M.,  {Bellagamba} F.,  {Radovich} M.,  {Roncarelli} M.,  {Sereno} M.,
  {Moscardini} L.,  {Bardelli} S.,   {Puddu} E.,  2019, \mn@doi [\mnras]
  {10.1093/mnras/stz294}, \href
  {https://ui.adsabs.harvard.edu/abs/2019MNRAS.485..498M} {485, 498}

\bibitem[\protect\citeauthoryear{{McClintock} et~al.,}{{McClintock}
  et~al.}{2019}]{mcclintock19}
{McClintock} T.,  et~al., 2019, \mn@doi [\mnras] {10.1093/mnras/sty2711}, \href
  {https://ui.adsabs.harvard.edu/abs/2019MNRAS.482.1352M} {482, 1352}

\bibitem[\protect\citeauthoryear{{Meneghetti}, {Rasia}, {Merten}, {Bellagamba},
  {Ettori}, {Mazzotta}, {Dolag}  \& {Marri}}{{Meneghetti}
  et~al.}{2010}]{meneghetti10}
{Meneghetti} M.,  {Rasia} E.,  {Merten} J.,  {Bellagamba} F.,  {Ettori} S.,
  {Mazzotta} P.,  {Dolag} K.,   {Marri} S.,  2010, \mn@doi [\aap]
  {10.1051/0004-6361/200913222}, \href
  {https://ui.adsabs.harvard.edu/abs/2010A&A...514A..93M} {514, A93}

\bibitem[\protect\citeauthoryear{{Meneghetti} et~al.,}{{Meneghetti}
  et~al.}{2014}]{meneghetti14}
{Meneghetti} M.,  et~al., 2014, \mn@doi [\apj] {10.1088/0004-637X/797/1/34},
  \href {https://ui.adsabs.harvard.edu/abs/2014ApJ...797...34M} {797, 34}

\bibitem[\protect\citeauthoryear{{Meneghetti} et~al.,}{{Meneghetti}
  et~al.}{2020}]{meneghetti20}
{Meneghetti} M.,  et~al., 2020, \mn@doi [Science] {10.1126/science.aax5164},
  \href {https://ui.adsabs.harvard.edu/abs/2020Sci...369.1347M} {369, 1347}

\bibitem[\protect\citeauthoryear{{Montes} \& {Trujillo}}{{Montes} \&
  {Trujillo}}{2019}]{montes19}
{Montes} M.,  {Trujillo} I.,  2019, \mn@doi [\mnras] {10.1093/mnras/sty2858},
  \href {https://ui.adsabs.harvard.edu/abs/2019MNRAS.482.2838M} {482, 2838}

\bibitem[\protect\citeauthoryear{{Mostoghiu}, {Knebe}, {Cui}, {Pearce},
  {Yepes}, {Power}, {Dave}  \& {Arth}}{{Mostoghiu} et~al.}{2019}]{mostoghiu19}
{Mostoghiu} R.,  {Knebe} A.,  {Cui} W.,  {Pearce} F.~R.,  {Yepes} G.,  {Power}
  C.,  {Dave} R.,   {Arth} A.,  2019, \mn@doi [\mnras] {10.1093/mnras/sty3306},
  \href {https://ui.adsabs.harvard.edu/abs/2019MNRAS.483.3390M} {483, 3390}

\bibitem[\protect\citeauthoryear{{Okabe}, {Nishimichi}, {Oguri}, {Peirani},
  {Kitayama}, {Sasaki}  \& {Suto}}{{Okabe} et~al.}{2018}]{okabe18}
{Okabe} T.,  {Nishimichi} T.,  {Oguri} M.,  {Peirani} S.,  {Kitayama} T.,
  {Sasaki} S.,   {Suto} Y.,  2018, \mn@doi [\mnras] {10.1093/mnras/sty1068},
  \href {https://ui.adsabs.harvard.edu/abs/2018MNRAS.478.1141O} {478, 1141}

\bibitem[\protect\citeauthoryear{{Okabe} et~al.,}{{Okabe}
  et~al.}{2020}]{okabe20}
{Okabe} T.,  et~al., 2020, \mn@doi [\mnras] {10.1093/mnras/staa1479}, \href
  {https://ui.adsabs.harvard.edu/abs/2020MNRAS.496.2591O} {496, 2591}

\bibitem[\protect\citeauthoryear{{Oogi} \& {Habe}}{{Oogi} \&
  {Habe}}{2013}]{oogi13}
{Oogi} T.,  {Habe} A.,  2013, \mn@doi [\mnras] {10.1093/mnras/sts047}, \href
  {https://ui.adsabs.harvard.edu/abs/2013MNRAS.428..641O} {428, 641}

\bibitem[\protect\citeauthoryear{{Peng}, {Ho}, {Impey}  \& {Rix}}{{Peng}
  et~al.}{2011}]{peng11}
{Peng} C.~Y.,  {Ho} L.~C.,  {Impey} C.~D.,   {Rix} H.-W.,  2011, {GALFIT:
  Detailed Structural Decomposition of Galaxy Images} (\mn@eprint {ascl}
  {1104.010})

\bibitem[\protect\citeauthoryear{{Planck Collaboration} et~al.,}{{Planck
  Collaboration} et~al.}{2016}]{planckcosmo16}
{Planck Collaboration} et~al., 2016, \mn@doi [\aap]
  {10.1051/0004-6361/201525830}, \href
  {https://ui.adsabs.harvard.edu/abs/2016A&A...594A..13P} {594, A13}

\bibitem[\protect\citeauthoryear{{Plazas}, {Meneghetti}, {Maturi}  \&
  {Rhodes}}{{Plazas} et~al.}{2019}]{plazas19}
{Plazas} A.~A.,  {Meneghetti} M.,  {Maturi} M.,   {Rhodes} J.,  2019, \mn@doi
  [\mnras] {10.1093/mnras/sty2737}, \href
  {https://ui.adsabs.harvard.edu/abs/2019MNRAS.482.2823P} {482, 2823}

\bibitem[\protect\citeauthoryear{Press, Teukolsky, Vetterling  \&
  Flannery}{Press et~al.}{1992}]{PresTeukVettFlan92}
Press W.~H.,  Teukolsky S.~A.,  Vetterling W.~T.,   Flannery B.~P.,  1992,
  Numerical Recipes in C, second edn.
Cambridge University Press, Cambridge, USA

\bibitem[\protect\citeauthoryear{{Privon} et~al.,}{{Privon}
  et~al.}{2020}]{privon20}
{Privon} G.~C.,  et~al., 2020, \mn@doi [\apj] {10.3847/1538-4357/ab8015}, \href
  {https://ui.adsabs.harvard.edu/abs/2020ApJ...893..149P} {893, 149}

\bibitem[\protect\citeauthoryear{{Ragone-Figueroa}, {Granato}, {Borgani}, {De
  Propris}, {Garc{\'\i}a Lambas}, {Murante}, {Rasia}  \&
  {West}}{{Ragone-Figueroa} et~al.}{2020}]{ragone20}
{Ragone-Figueroa} C.,  {Granato} G.~L.,  {Borgani} S.,  {De Propris} R.,
  {Garc{\'\i}a Lambas} D.,  {Murante} G.,  {Rasia} E.,   {West} M.,  2020,
  \mn@doi [\mnras] {10.1093/mnras/staa1389}, \href
  {https://ui.adsabs.harvard.edu/abs/2020MNRAS.495.2436R} {495, 2436}

\bibitem[\protect\citeauthoryear{{Rasia} et~al.,}{{Rasia}
  et~al.}{2015}]{rasia15}
{Rasia} E.,  et~al., 2015, \mn@doi [\apjl] {10.1088/2041-8205/813/1/L17}, \href
  {https://ui.adsabs.harvard.edu/abs/2015ApJ...813L..17R} {813, L17}

\bibitem[\protect\citeauthoryear{{Rykoff} et~al.,}{{Rykoff}
  et~al.}{2016}]{rykoff16}
{Rykoff} E.~S.,  et~al., 2016, \mn@doi [\apjs] {10.3847/0067-0049/224/1/1},
  \href {https://ui.adsabs.harvard.edu/abs/2016ApJS..224....1R} {224, 1}

\bibitem[\protect\citeauthoryear{{Sartoris} et~al.,}{{Sartoris}
  et~al.}{2016}]{sartoris16}
{Sartoris} B.,  et~al., 2016, \mn@doi [\mnras] {10.1093/mnras/stw630}, \href
  {https://ui.adsabs.harvard.edu/abs/2016MNRAS.459.1764S} {459, 1764}

\bibitem[\protect\citeauthoryear{{Schrabback} et~al.,}{{Schrabback}
  et~al.}{2020}]{schrabback20}
{Schrabback} T.,  et~al., 2020, arXiv e-prints, \href
  {https://ui.adsabs.harvard.edu/abs/2020arXiv200907591S} {p. arXiv:2009.07591}

\bibitem[\protect\citeauthoryear{{Sheth} \& {Tormen}}{{Sheth} \&
  {Tormen}}{1999}]{sheth99}
{Sheth} R.~K.,  {Tormen} G.,  1999, \mn@doi [\mnras]
  {10.1046/j.1365-8711.1999.02692.x}, \href
  {https://ui.adsabs.harvard.edu/abs/1999MNRAS.308..119S} {308, 119}

\bibitem[\protect\citeauthoryear{{Shi}, {Osato}, {Kurita}  \& {Takada}}{{Shi}
  et~al.}{2021}]{shi21}
{Shi} J.,  {Osato} K.,  {Kurita} T.,   {Takada} M.,  2021, arXiv e-prints,
  \href {https://ui.adsabs.harvard.edu/abs/2021arXiv210412329S} {p.
  arXiv:2104.12329}

\bibitem[\protect\citeauthoryear{{Shirasaki}, {Nagai}  \& {Lau}}{{Shirasaki}
  et~al.}{2016}]{shirasaki16}
{Shirasaki} M.,  {Nagai} D.,   {Lau} E.~T.,  2016, \mn@doi [\mnras]
  {10.1093/mnras/stw1263}, \href
  {https://ui.adsabs.harvard.edu/abs/2016MNRAS.460.3913S} {460, 3913}

\bibitem[\protect\citeauthoryear{{Sunayama} et~al.,}{{Sunayama}
  et~al.}{2020}]{sunayama20}
{Sunayama} T.,  et~al., 2020, \mn@doi [\mnras] {10.1093/mnras/staa1646}, \href
  {https://ui.adsabs.harvard.edu/abs/2020MNRAS.496.4468S} {496, 4468}

\bibitem[\protect\citeauthoryear{{Tenneti}, {Mandelbaum}, {Di Matteo},
  {Kiessling}  \& {Khandai}}{{Tenneti} et~al.}{2015}]{tenneti15}
{Tenneti} A.,  {Mandelbaum} R.,  {Di Matteo} T.,  {Kiessling} A.,   {Khandai}
  N.,  2015, \mn@doi [\mnras] {10.1093/mnras/stv1625}, \href
  {https://ui.adsabs.harvard.edu/abs/2015MNRAS.453..469T} {453, 469}

\bibitem[\protect\citeauthoryear{{To} et~al.,}{{To} et~al.}{2021}]{to21}
{To} C.,  et~al., 2021, \mn@doi [\prl] {10.1103/PhysRevLett.126.141301}, \href
  {https://ui.adsabs.harvard.edu/abs/2021PhRvL.126n1301T} {126, 141301}

\bibitem[\protect\citeauthoryear{Van~der Walt, Sch{\"o}nberger, Nunez-Iglesias,
  Boulogne, Warner, Yager, Gouillart  \& Yu}{Van~der Walt
  et~al.}{2014}]{walt14}
Van~der Walt S.,  Sch{\"o}nberger J.~L.,  Nunez-Iglesias J.,  Boulogne F.,
  Warner J.~D.,  Yager N.,  Gouillart E.,   Yu T.,  2014, PeerJ, 2, e453

\bibitem[\protect\citeauthoryear{{Velliscig} et~al.,}{{Velliscig}
  et~al.}{2015}]{velliscig15}
{Velliscig} M.,  et~al., 2015, \mn@doi [\mnras] {10.1093/mnras/stv1690}, \href
  {https://ui.adsabs.harvard.edu/abs/2015MNRAS.453..721V} {453, 721}

\bibitem[\protect\citeauthoryear{{Vikhlinin} et~al.,}{{Vikhlinin}
  et~al.}{2009a}]{vikhlinin09a}
{Vikhlinin} A.,  et~al., 2009a, \mn@doi [\apj] {10.1088/0004-637X/692/2/1033},
  \href {https://ui.adsabs.harvard.edu/abs/2009ApJ...692.1033V} {692, 1033}

\bibitem[\protect\citeauthoryear{{Vikhlinin} et~al.,}{{Vikhlinin}
  et~al.}{2009b}]{vikhlinin09}
{Vikhlinin} A.,  et~al., 2009b, \mn@doi [\apj] {10.1088/0004-637X/692/2/1060},
  \href {https://ui.adsabs.harvard.edu/abs/2009ApJ...692.1060V} {692, 1060}

\bibitem[\protect\citeauthoryear{{Wang} et~al.,}{{Wang} et~al.}{2018}]{wang18}
{Wang} Y.,  et~al., 2018, \mn@doi [\apj] {10.3847/1538-4357/aae52e}, \href
  {https://ui.adsabs.harvard.edu/abs/2018ApJ...868..130W} {868, 130}

\bibitem[\protect\citeauthoryear{{Wittman}, {Foote}  \& {Golovich}}{{Wittman}
  et~al.}{2019}]{wittman19}
{Wittman} D.,  {Foote} D.,   {Golovich} N.,  2019, \mn@doi [\apj]
  {10.3847/1538-4357/ab0a0a}, \href
  {https://ui.adsabs.harvard.edu/abs/2019ApJ...874...84W} {874, 84}

\bibitem[\protect\citeauthoryear{{Zhang} \& {Annis}}{{Zhang} \&
  {Annis}}{2022}]{zhangannis22}
{Zhang} Y.,  {Annis} J.,  2022, arXiv e-prints, \href
  {https://ui.adsabs.harvard.edu/abs/2022arXiv220102167Z} {p. arXiv:2201.02167}

\bibitem[\protect\citeauthoryear{{Zhang} et~al.,}{{Zhang}
  et~al.}{2019a}]{zhang19b}
{Zhang} Y.,  et~al., 2019a, \mn@doi [\mnras] {10.1093/mnras/stz1361}, \href
  {https://ui.adsabs.harvard.edu/abs/2019MNRAS.487.2578Z} {487, 2578}

\bibitem[\protect\citeauthoryear{{Zhang} et~al.,}{{Zhang}
  et~al.}{2019b}]{zhang19}
{Zhang} Y.,  et~al., 2019b, \mn@doi [\apj] {10.3847/1538-4357/ab0dfd}, \href
  {https://ui.adsabs.harvard.edu/abs/2019ApJ...874..165Z} {874, 165}

\bibitem[\protect\citeauthoryear{{von der Linden} et~al.,}{{von der Linden}
  et~al.}{2014}]{vdlinden14}
{von der Linden} A.,  et~al., 2014, \mn@doi [\mnras] {10.1093/mnras/stu1423},
  \href {https://ui.adsabs.harvard.edu/abs/2014MNRAS.443.1973V} {443, 1973}

\makeatother
\end{thebibliography}












\label{lastpage}
\end{document}